\documentclass[journal,twocolumn,twoside]{IEEEtranTCOM}
\usepackage[pdftex]{graphicx}
\usepackage[cmex10]{amsmath}
\usepackage{amssymb}

\usepackage{caption}
\usepackage{subcaption}
\usepackage{cite}
\usepackage{psfrag}
\usepackage{epstopdf}
\usepackage{algorithm}
\usepackage{algorithmic}
\usepackage{widetext}
\usepackage{float}
\usepackage{color}
\usepackage{soul}
\usepackage{upgreek}
\usepackage{multirow}
\usepackage[justification=centerlast]{caption}
\usepackage{tikz,pgfplots}
\usepackage{dblfloatfix}
\usepackage{adjustbox}
\DeclareUnicodeCharacter{2005}{\hspace{0.01em}}
\usetikzlibrary{calc}
\usepackage{mathtools}
\usetikzlibrary{arrows}
\pgfplotsset{compat=newest}
\usepgfplotslibrary{fillbetween}

\definecolor{myBlue}{RGB}{72,125,215}
\definecolor{myOrange}{RGB}{118,54,45}
\definecolor{InfinBlue}{RGB}{72,72,51}
\definecolor{bblue}{HTML}{4F81BD}
\definecolor{rred}{HTML}{C0504D}
\definecolor{ggreen}{HTML}{9BBB59}
\definecolor{ppurple}{HTML}{9F4C7C}

\usetikzlibrary{patterns}

\ifCLASSINFOpdf
\else
\fi
\begin{document}

\title{Deep Neural Network-aided Soft-Demapping in Coherent Optical Systems: Regression versus Classification}

     \pgfplotsset{
        compat=1.3, 
        my axis style/.style={
            every axis plot post/.style={/pgf/number format/fixed},
            ybar=5pt,
            bar width=8pt,
            x=1.7cm,
            axis on top,
            enlarge x limits=0.1,
            symbolic x coords={MLP, biLSTM, ESN, CNN+MLP, CNN+biLSTM},
            visualization depends on=rawy\as\rawy, 
            nodes near coords={%
                \pgfmathprintnumber[precision=2]{\rawy}
            },
            every node near coord/.append style={rotate=90, anchor=west},
            tick label style={font=\footnotesize},
            xtick distance=1,
        },
    }
%

\author{Pedro J. Freire, Jaroslaw E. Prilepsky, Yevhenii Osadchuk,  Sergei K. Turitsyn, Vahid Aref
\thanks{This paper was supported by the EU  Horizon 2020 program under the Marie Sklodowska-Curie grant agreement 813144 (REAL-NET). JEP is supported by Leverhulme Trust, Grant No. RP-2018-063. SKT acknowledges support of the EPSRC project TRANSNET. Additionally, The authors further thank the associate editor and unanimous reviewers for insightful discussions  and suggestions.}
\thanks{Pedro J. Freire, Yevhenii Osadchuk, Jaroslaw E. Prilepsky  and Sergei K. Turitsyn are with Aston Institute of Photonic Technologies, Aston University, United Kingdom, p.freiredecarvalhosouza@aston.ac.uk.}
\thanks{Vahid Aref is with Nokia Bell Labs, Lorenzstraße 10, 70435, Stuttgart, Germany, vahid.aref@nokia.com.}

\thanks{Manuscript received xxx 19, zzz; revised January 11, yyy.}}

\maketitle
\begin{abstract}
We examine here what type of predictive modelling, classification, or regression, using neural networks (NN), fits better the task of soft-demapping based post-processing in coherent optical communications, where the transmission channel is nonlinear and dispersive. For the first time, we present possible drawbacks in using each type of predictive task in a machine learning context, considering the nonlinear coherent optical channel equalization/soft-demapping problem. We study two types of equalizers based on the feed-forward and recurrent NNs, for several transmission scenarios, in linear and nonlinear regimes of the optical channel. We point out that even though from the information theory perspective the cross-entropy loss (classification) is the most suitable option for our problem, the NN models based on the cross-entropy loss function can severely suffer from learning problems. The latter translates into the fact that regression-based learning is typically superior in terms of delivering higher Q-factor and achievable information rates. In short, we show by empirical evidence that loss functions based on cross-entropy may not be necessarily the most suitable option for training communication systems in practical scenarios when overfitting- and vanishing gradients-related problems come into play. 
\end{abstract}


\IEEEpeerreviewmaketitle

\section{Introduction}

\IEEEPARstart{T}o improve the performance of optical fiber systems, it is important to mitigate the detrimental impact of linear and, most importantly, nonlinear transmission impairments that cape the systems' throughput\cite{akc16,winzer2018fiber,Cartledge:17}. Numerous digital signal processing (DSP) algorithms have been proposed and studied for the optical fiber channel equalization problem~\cite{Cartledge:17}. Over the past few years, the ``conventional'' equalizers/soft-demappers have started to evolve toward the designs incorporating machine learning techniques \cite{Khan:17, Zibar01,Zibar02}. In particular, the neural network (NN) based channel equalization/demapping has recently become a topic of intensive research in optical communications, due to its capability in mitigation of linear and nonlinear impairments in optical channels and transceivers. For instance, in Refs.~\cite{6975096, WANG20171, zhang2019field, freire2020complex} the authors successfully applied artificial NN-based nonlinear equalizers for impairment compensation in coherent transmission links, while in Refs.~\cite{hager2018nonlinear, Bitachon20}, more advanced deep learning algorithms were introduced and compared with the performance rendered by conventional DSP methods. Additionally, in Ref.~\cite{schaedlerrecurrent} the authors suggested a bit-wise soft-demapper model based on bidirectional recurrent NN applied for nonlinear ISI compensation in coherent links. Recurrent NNs in the form of long short-term memory (LSTM) has shown great potential in nonlinearities mitigation~\cite{9184798,freire2021performance}. Additionally, the end-to-end learning designs \cite{Hoydis} that deal with coherent constellation optimization have been intensively studied in the literature \cite{Karanov2020,Karanov2018,zhu2019autoencoder, Alvarado2020, aoudia2020end,neskorniuk2021end,aref2021end}: in these works, the end-to-end system is typically composed as an auto-encoder trained with the cross-entropy loss (CEL) function. Finally, we notice that the authors of Ref.~\cite{end_to_end_MSE} recently investigated the back-propagation blocking problem in training the end-to-end NN designs, showing the possible problems in using the CEL. In the same Ref.~\cite{end_to_end_MSE}, a version of the mean squared error (MSE) loss, for the end-to-end systems, was proposed, where the resulting performance was better than the one delivered by the CEL-based NN models. 

The most popular and, perhaps, simple approach to improving the channel capacity is the receiver-based equalizer -- a special-purpose DSP device that can (partially) reverse the distortions incurred by a signal when passing over the optical channel. These equalizers are typically designed and optimized based on the minimum-mean-squared-error (MMSE) criteria.
The usefulness of MMSE equalizers is stipulated for several reasons, including: (i) the MMSE is quite convenient for mathematical optimization because of convexity and differentiability of its objective function; and (ii) the MMSE equalizer is usually optimized independently of the underlying waveform or modulation format. In this paper, we call this category of NN models based on MMSE as a regression equalizer (Reg.), but we can term them as a dual-stage soft-demapping (i.e. the NN MMSE post-equalizer followed by the AWGN demapper) \cite{Georg2021book}.

\begin{figure*}[!ht]
    \centering
    \includegraphics[width=0.8\linewidth]{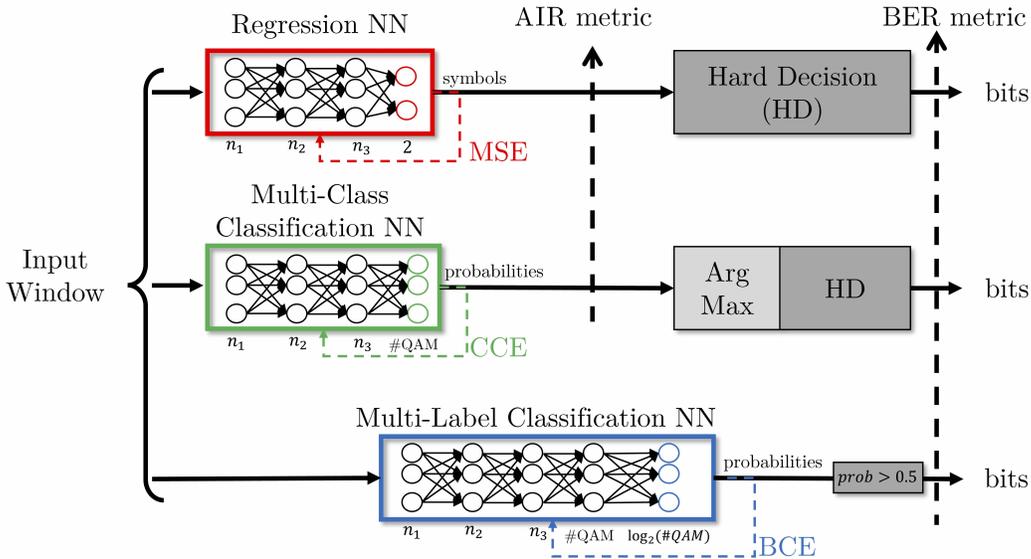}
    \caption{Scheme of the three classic configurations of NN models, using the MLP as an exemplary NN core: regression (top), multi-class classification (middle), and multi-label classification (bottom), in the context of channel equalization and soft demapping in communications.}
    \label{fig:schemes_NN}
\end{figure*}

Another approach is to design a model incorporating the decision step, which corresponds to the classifier in the machine learning literature, see, e.g., \cite[Chapter 6]{Goodfellowbook2016}. Such a device is convenient insofar as the transmitted signal in digital communications is usually generated from a discrete finite-size constellation, e.g. quadrature amplitude modulation, QAM. This approach has received more attention recently \cite{Hoydis,dorner2017deep,deligiannidis2020compensation, schaedlerrecurrent, Georg,aref2021end}, in view of the following reasons: (i) the classifier is optimized for the specific in-use modulation format; (ii) it directly maximizes the information rate, the main objective of the channel equalization, and outputs the likelihoods for each received symbol, a more suitable metric for the subsequent forward error correction; (iii) and, even more importantly, it can adapt itself to the correct statistical channel characteristics. In our paper, we call this category of predictive model a multi-class classifier (MC Class.), and in our communication study case, we can see them as single-stage soft demapping, in which we use the categorical CEL in the learning process.

Here, we also mention that another possible approach allows us to build on the bitwise representation of the received symbols and to train the NN model using the binary cross-entropy loss (BCEL) as a one-step setup \cite{Georg}. This type of receiver model, named in machine learning as a \textit{multi-label classifier} (MB Class.), is used in end-to-end learning~\cite{end_to_end_MSE,aref2021end, Karanov2020} or for the soft-demapping in short-reach transmission scenarios~\cite{Georg}. In our current work, we mostly focus on the multi-class classification and the regression approaches, but in the last part of the results section, we additionally perform a comparative study for the multi-label classification, to make sure that the observed effects refer to this case as well. The NN models for the regression equalizer, multi-class classifier, and the multi-label classifier used in this paper are schematically given in Fig.\ref{fig:schemes_NN}\footnote{To get the bits out of the probabilities in the multi-label classification case, we use a simple decision block that returns 1 if the probability was higher than 0.5 and  0 otherwise.}.

It is worth mentioning at this point that the transfer of the methods developed in the field of machine learning to optical communications should be taken into account \textit{the underlying peculiarities and challenges of NN algorithms themselves}, while those are often overlooked when designing the equalizers/soft-demappers in communications-related literature. From Ref.\cite{Georg2021book} we infer that cross-entropy is the most suitable loss function for communication applications, accounting for its meaning in information theory. However, in our paper, we intend to increase the researcher's awareness of possible training problems that we observed when dealing with the CEL gradient-descent-based learning. Our case study here specifically refers to the complex problem of symbols demapping in long-haul coherent optical transmission, in which the memory from dispersion and fiber nonlinearity plays an important role; however, this may have less relevance to the short-haul/back-to-back scenarios, which, perhaps, deserve a separate study.

A fair comparison between the regression and classification is quite challenging, as 
these two produce different output variable types: discrete versus continuous, respectively.
Such comparison studies have been carried out in theoretical machine learning-related works, where the output of the classification was analyzed with different loss functions \cite{golik2013cross, nar2018cross, bosman2020visualising}. However, to our knowledge,  only in Ref.~\cite{lathuiliere2019comprehensive} directly explained the motivation to solve problems with regression instead of classification for specific tasks. In Refs.~\cite{li2016online,mukherjee2021joint, liu2018joint, liu2017deep, wu2021joint,chen2019joint}, it was recognized that both regression and classification tasks have potential downsides: the regression model cannot utilize the full flexibility of discriminative NN models (meaning that the statistic of the regression model is assumed to be Gaussian, while the classification model does not depend on the type of statistics, and, thus, is more flexible.); the classification model, in turn, is not able to capture misclassification differences when we misinterpret the classes, which can degrade the modelling quality. Thus, in these works, the combination of regression and classification was proposed as the best problem fit, but we do not address it here. 

The interest in finding the best solution between the regression- and classification-based NN models, frequently arising in different areas, prompts us to conduct the investigation of this dilemma for the development of NN-based soft-demappers in coherent optical systems. For the optical channel demapping, this comparison may be made more evident by contrasting the multi-class classifier output to that obtained with the regression, in terms of bit error rate (BER) (i.e. using a hard decision metric) or with respect to the achievable information rate (AIR), i.e. using a soft decision metric. In this paper, we compare the performance of the multi-class CEL classifier and regression MMSE predictive models and expose the potential drawbacks of each task specifically for the NN-based long-haul optical channel soft-demapping problem.
Our findings and rationale behind the observed interplay between the regression and classification results can be briefly summarized as follows.
\begin{itemize}
    \item For the MSE training process we recommend using an \textit{early stopping at the point where the achievable information rate is maximal}, but not at the point of minimal MSE, as to avoid underestimation in the calculation of the achievable information rate and ``jail-window'' constellation patterning \cite{freire2022neural}. 
    \item The CEL landscape is prone to sharp local minima, revealing large gradients at the points where the training loss value is close to zero. Thus, we arrive at the overfitting of the model, where the loss value for the training is much lower than that for the testing dataset. In contrast, the MSE produces wide local minima, where we have relatively small gradients in the vicinity of the training loss minimum. It allows us to avoid overfitting more efficiently. 
    \item The CEL was incapable of delivering sufficient training quality due to the vanishing of  back-propagating gradients. In contrast to regression (with the MSE loss function), classification could not handle well \textit{the high-accuracy systems} (with a very low number of errors in the training datasets) that we typically have in optical communication applications, causing the NN to rapidly converge to a local minimum, where the gradient is close to zero.  
 \item As it was also observed in computer vision tasks \cite{lin2017focal}, the CEL function in our case gets overwhelmed by the ``easy prediction'', which is typical for the high accuracy systems. 
\end{itemize}

These effects, and their consequence on the NN's learning quality and performance, are considered in detail further. As a result, it is exactly the classifiers' training challenges due to the aforementioned reasons, but not the optimally/suboptimality of the loss function for the demapping, that plays a crucial role in defining the performance of NN-based models. This conclusion can be somewhat counter-intuitive in view of the information theory arguments based on the optimality of the demapper \cite{Georg2021book}. we also point out that our findings here coincide with those presented in very recent Ref.~\cite{end_to_end_MSE} and Ref.~\cite{bosman2020visualising}, allowing us to conclude that the observed effects can pertain not only to the NN-based equalization/demapping but also to the other types of NN-based systems (e.g. to end-to-end learning).

\section{Deep Neural Network-aided Soft-Demapping}

\begin{figure*}[ht!]    
\begin{subfigure}[t]{0.5\textwidth}
            \centering
                \begin{tikzpicture}[scale=0.9]
\begin{axis}[
xmin=0,xmax=100,
axis y line*=left,
xlabel= Training Epochs,
ylabel style = {align=center},
ylabel={AIR [bits/symbol] \ref{pgfplots:plot1}
},
grid=both,   
grid style={dashed},  
]

\addplot[color=red, mark=square, very thick]
    coordinates {
(0,3.48564200000000013801582)
(1,3.49564200000000013801582)
(2,3.53452799999999989211119)
(3,3.55098400000000014031798)
(4,3.55376100000000016976287)
(5,3.56341900000000011417001)
(6,3.56613500000000005485390)
(7,3.56619300000000016837021)
(8,3.56557700000000021844926)
(9,3.56158200000000002560796)
(10,3.56176499999999984780175)
(11,3.55766699999999991277377)
(12,3.55693200000000020466473)
(13,3.56013199999999985223553)
(14,3.55990999999999990777155)
(15,3.55580600000000002225420)
(16,3.54486500000000015475621)
(17,3.53880099999999986337684)
(18,3.54142700000000010263079)
(19,3.52557700000000018292212)
(20,3.49820200000000003370815)
(21,3.48488899999999990342303)
(22,3.45964200000000010604140)
(23,3.43558399999999997120881)
(24,3.38443799999999983540988)
(25,3.32544299999999992678568)
(26,3.27618099999999978777510)
(27,3.23991400000000018266633)
(28,3.19442899999999996296651)
(29,3.17039099999999995915800)
(30,3.16549099999999983268140)
(31,3.14109199999999999519673)
(32,3.13483899999999993113420)
(33,3.11744799999999999684519)
(34,3.12155799999999983285193)
(35,3.11156999999999994699351)
(36,3.11036700000000010390977)
(37,3.10287099999999993471533)
(38,3.09840999999999988645527)
(39,3.09405300000000016424906)
(40,3.09067500000000006110668)
(41,3.08896600000000010055601)
(42,3.08874499999999985178079)
(43,3.08380000000000009663381)
(44,3.08417500000000011084467)
(45,3.09000299999999983313614)
(46,3.07859699999999980590815)
(47,3.07547699999999979425525)
(48,3.07597100000000001074341)
(49,3.07943500000000014438228)
(50,3.07316799999999989978505)
(51,3.06965200000000004720846)
(52,3.06676100000000007028689)
(53,3.06584199999999995611688)
(54,3.06321600000000016095214)
(55,3.06241500000000010928147)
(56,3.06331199999999981287147)
(57,3.06089199999999994616928)
(58,3.06057699999999988094146)
(59,3.06140899999999982483700)
(60,3.06456799999999995876010)
(61,3.06227800000000005553602)
(62,3.05796799999999979746690)
(63,3.05595900000000009200107)
(64,3.05501699999999987156230)
(65,3.06085299999999982389909)
(66,3.05345900000000014529178)
(67,3.05052300000000009561063)
(68,3.05253199999999980107646)
(69,3.05871899999999996566658)
(70,3.05451300000000003365130)
(71,3.05827699999999991220534)
(72,3.04930799999999990745891)
(73,3.05034500000000008412826)
(74,3.05234600000000000363798)
(75,3.04964099999999982415488)
(76,3.04861100000000018184210)
(77,3.04509799999999986042098)
(78,3.04753799999999985814725)
(79,3.04465699999999994673772)
(80,3.04628299999999985203658)
(81,3.05106900000000003103651)
(82,3.04563400000000017442403)
(83,3.04668200000000011229417)
(84,3.04803399999999991010213)
(85,3.05098199999999986076205)
(86,3.04895100000000018880542)
(87,3.04184099999999979502263)
(88,3.04537300000000010768986)
(89,3.04441899999999998627231)
(90,3.05099199999999992627409)
(91,3.04161599999999987531396)
(92,3.04383399999999992857624)
(93,3.04160999999999992482458)
(94,3.04226300000000016154900)
(95,3.04517300000000012971668)
(96,3.03868200000000010518875)
(97,3.04122499999999984510168)
(98,3.04377100000000000434852)
(99,3.04104600000000013793056)
(100,3.04330500000000014892976)
    };

\label{pgfplots:plot1}
\end{axis}
\begin{axis}[
xmin=0,xmax=100,
axis y line*=right,
axis x line=none,
ylabel style = {align=center},
ylabel={Mean Squared Error \ref{pgfplots:plot2}},        
]
\addplot[color=blue, mark=*, very thick]
    coordinates {
(0,0.03335899999999999798739)
(1,0.03135899999999999798739)
(2,0.02965100000000000027289)
(3,0.02885499999999999870548)
(4,0.02842300000000000034794)
(5,0.02777800000000000060774)
(6,0.02742599999999999899059)
(7,0.02722099999999999866973)
(8,0.02710100000000000008971)
(9,0.02709500000000000102807)
(10,0.02699100000000000110223)
(11,0.02690700000000000036149)
(12,0.02672199999999999922573)
(13,0.02655499999999999874434)
(14,0.02607200000000000142397)
(15,0.02560399999999999828826)
(16,0.02514200000000000115308)
(17,0.02453800000000000078315)
(18,0.02399499999999999896860)
(19,0.02321399999999999852474)
(20,0.02300900000000000167333)
(21,0.02204400000000000109379)
(22,0.02157500000000000042744)
(23,0.02080200000000000104539)
(24,0.01982199999999999934230)
(25,0.01887399999999999841704)
(26,0.01851300000000000167688)
(27,0.01831500000000000141775)
(28,0.01775600000000000094902)
(29,0.01756399999999999975153)
(30,0.01777199999999999960321)
(31,0.01735899999999999943068)
(32,0.01750300000000000119615)
(33,0.01724099999999999938138)
(34,0.01734799999999999883804)
(35,0.01728000000000000022093)
(36,0.01735799999999999843059)
(37,0.01720899999999999860356)
(38,0.01725099999999999897393)
(39,0.01721699999999999966538)
(40,0.01715200000000000057909)
(41,0.01720100000000000101119)
(42,0.01719899999999999901101)
(43,0.01711000000000000020872)
(44,0.01712699999999999986300)
(45,0.01736800000000000149258)
(46,0.01707300000000000136935)
(47,0.01716300000000000117173)
(48,0.01706600000000000130762)
(49,0.01717700000000000129519)
(50,0.01707799999999999943090)
(51,0.01708099999999999896172)
(52,0.01707600000000000090017)
(53,0.01707499999999999990008)
(54,0.01707799999999999943090)
(55,0.01704699999999999965317)
(56,0.01705200000000000118416)
(57,0.01701400000000000134470)
(58,0.01703999999999999959144)
(59,0.01706999999999999836908)
(60,0.01710499999999999867772)
(61,0.01707399999999999889999)
(62,0.01703899999999999859135)
(63,0.01702899999999999899880)
(64,0.01700700000000000128297)
(65,0.01707600000000000090017)
(66,0.01702800000000000146816)
(67,0.01706200000000000077671)
(68,0.01703400000000000052980)
(69,0.01708499999999999949263)
(70,0.01714099999999999998646)
(71,0.01711800000000000127054)
(72,0.01703999999999999959144)
(73,0.01701000000000000081379)
(74,0.01707200000000000036926)
(75,0.01700899999999999981370)
(76,0.01699200000000000015943)
(77,0.01698300000000000156697)
(78,0.01701499999999999887534)
(79,0.01697999999999999856670)
(80,0.01698699999999999862843)
(81,0.01704299999999999912226)
(82,0.01698200000000000056688)
(83,0.01698600000000000109779)
(84,0.01700199999999999975198)
(85,0.01712299999999999933209)
(86,0.01700099999999999875189)
(87,0.01694899999999999878897)
(88,0.01699000000000000162870)
(89,0.01700000000000000122125)
(90,0.01709099999999999855427)
(91,0.01697099999999999997424)
(92,0.01702400000000000093725)
(93,0.01701000000000000081379)
(94,0.01697299999999999850497)
(95,0.01698200000000000056688)
(96,0.01697600000000000150524)
(97,0.01698699999999999862843)
(98,0.01702100000000000140643)
(99,0.01698200000000000056688)
(100,0.01699600000000000069034)
    };

\label{pgfplots:plot2}
\end{axis}
    \node at (3,4.5) {\includegraphics[width=1.9cm]{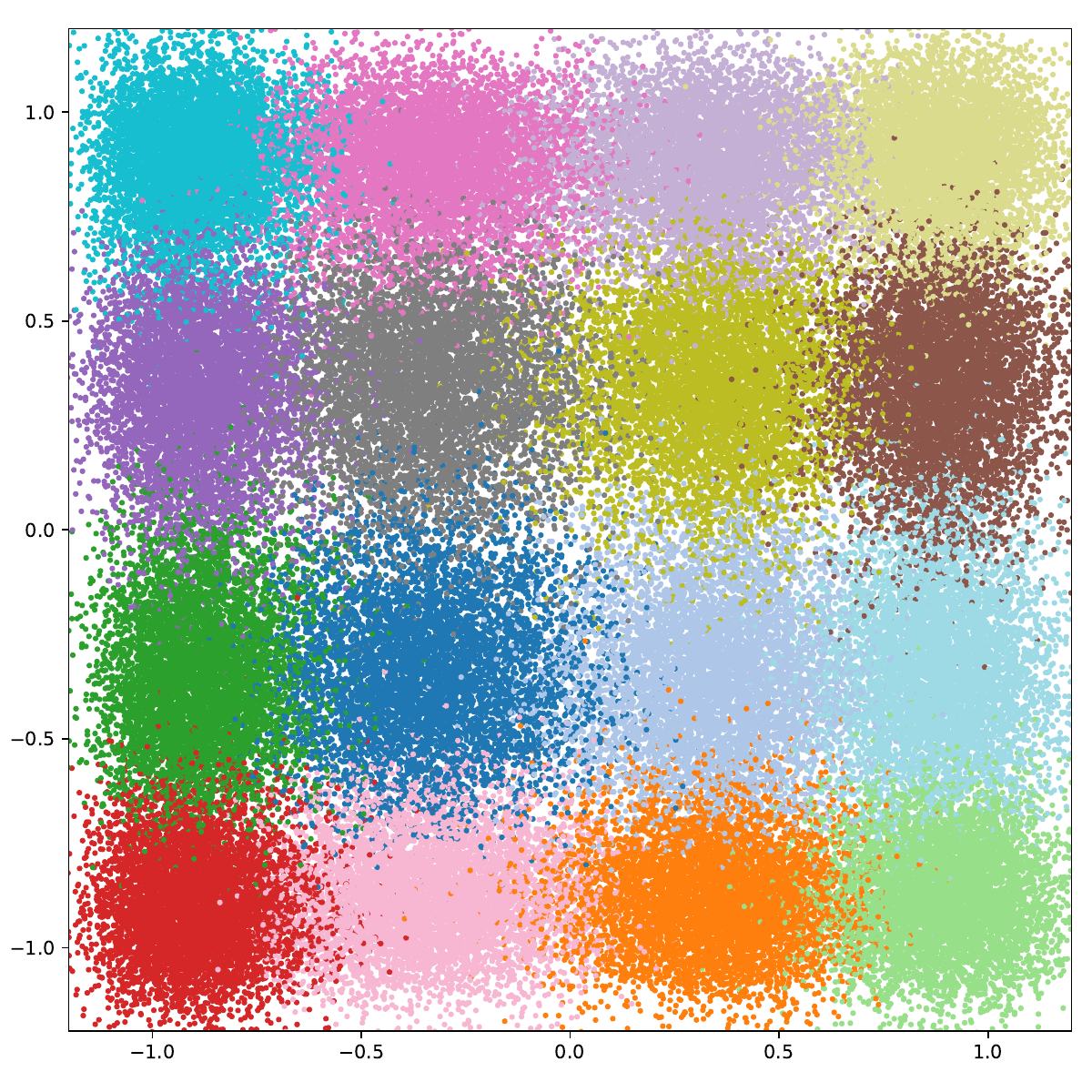}};
    \node at (5,2.2) {\includegraphics[width=1.9cm]{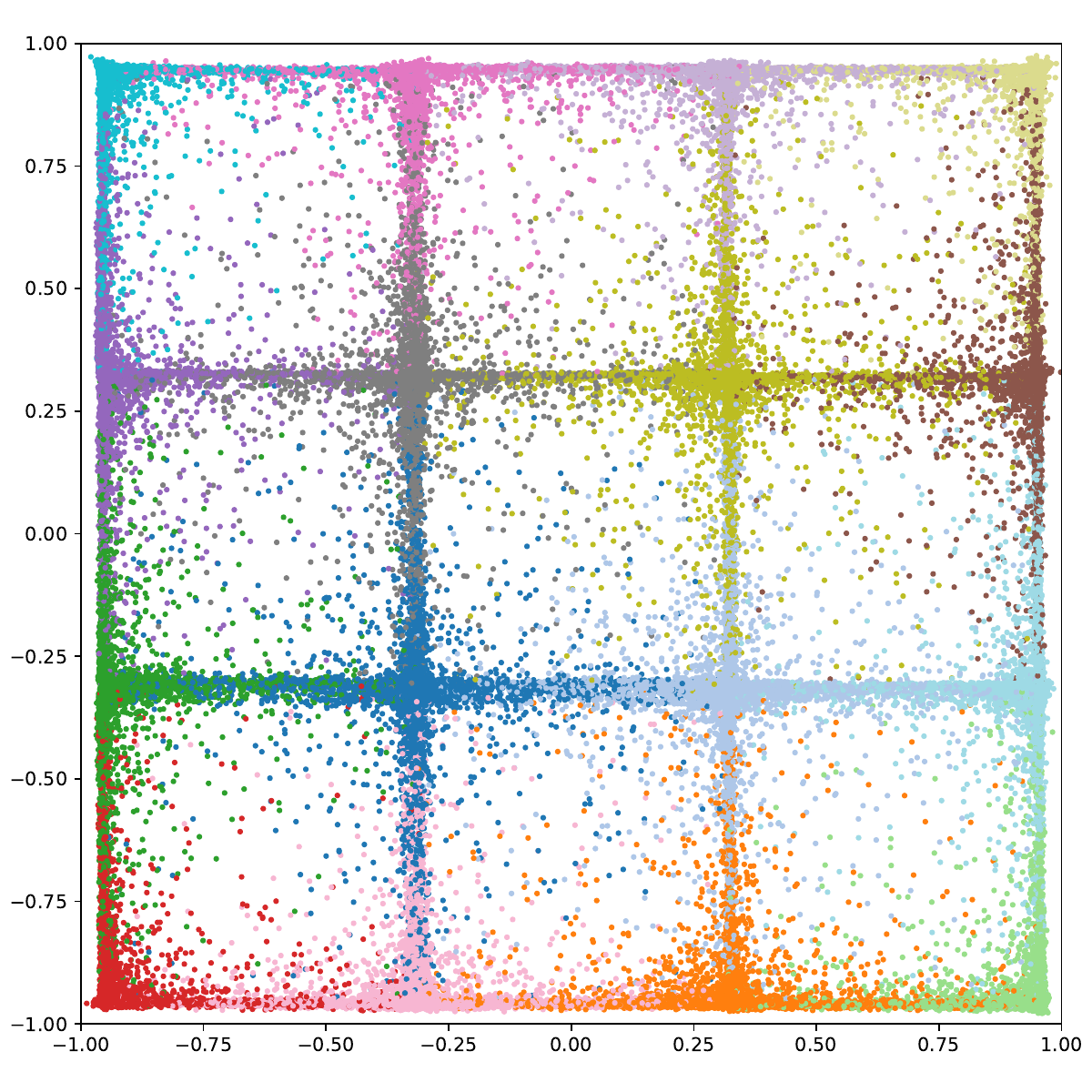}};
\node (mark) [draw, red, circle, minimum size = 6pt] at (0.45, 5.2){};
    \draw [-stealth,red](0.45, 5.2) -- (2.1, 4.5);
\node (mark) [draw, blue, circle, minimum size = 6pt] at (5.92, 0.5){};
    \draw [-stealth,blue](5.92, 0.5) -- (5, 1.2);

    \end{tikzpicture}
    \caption{}
\end{subfigure}
\hspace{\fill}
\begin{subfigure}[t]{0.5\textwidth}
            \centering
                \begin{tikzpicture}[scale=0.9]
      \begin{axis} [
        xlabel={$E_b/N_0$ [dB]},
        ylabel={AIR [bits/symbol]},
        grid=both, 
        ymax=4.5,
    	xmin=2, xmax=29,
        legend style={legend pos=south west, legend cell align=left,fill=white, fill opacity=0.6, draw opacity=1,text opacity=1},
    	grid style={dashed}]
        ]
              \addplot[color=red, mark=square, very thick]
    coordinates {
(2,1.30330499999999993576694)
(3,1.50483999999999995544897)
(4,1.72193299999999993588062)
(5,1.94424200000000002575007)
(6,2.17260300000000006193090)
(7,2.41433499999999989782395)
(8,2.65203299999999986269472)
(9,2.90317399999999992132871)
(10,3.14527299999999998547651)
(11,3.37191000000000018488322)
(12,3.56570400000000020668267)
(13,3.72390800000000021796609)
(14,3.84376700000000015577939)
(15,3.92213799999999990220090)
(16,3.96814899999999992630251)
(17,3.98832100000000000505906)
(18,3.99706800000000006534151)
(19,3.99926799999999982304644)
(20,3.99994000000000005101697)
(21,3.99998299999999984422061)
(22,4.00000000000000000000000)
(23,4.00000000000000000000000)
(24,4.00000000000000000000000)
(25,4.00000000000000000000000)
(26,4.00000000000000000000000)
(27,4.00000000000000000000000)
(28,4.00000000000000000000000)
(29,4.00000000000000000000000)
    };

    \addlegendentry{\footnotesize{Regression model with AIR early stopping}};
    
              \addplot[color=blue, mark=o, very thick]
    coordinates {
(2,1.20205399999999995586109)
(3,1.40050000000000007815970)
(4,1.61019599999999996065014)
(5,1.82152500000000006075140)
(6,2.03444599999999997663735)
(7,2.24474999999999980104803)
(8,2.44102599999999991808863)
(9,2.62518699999999993721644)
(10,2.77643599999999990401989)
(11,2.91079299999999996373390)
(12,3.04265900000000000247269)
(13,3.18615599999999998814815)
(14,3.34415899999999988168042)
(15,3.50743800000000005567813)
(16,3.67694799999999988315835)
(17,3.81872899999999981801579)
(18,3.93857599999999985485033)
(19,3.97793999999999980943244)
(20,3.99572700000000002873435)
(21,4.00000000000000000000000)
(22,4.00000000000000000000000)
(23,4.00000000000000000000000)
(24,4.00000000000000000000000)
(25,4.00000000000000000000000)
(26,4.00000000000000000000000)
(27,4.00000000000000000000000)
(28,4.00000000000000000000000)
(29,4.00000000000000000000000)
    };
    \addlegendentry{\footnotesize{Regression model with MSE minimization}};

    \addplot[color=cyan, mark=+,dashed, very thick]        coordinates {
(2,1.34521295754624770069086)
(3,1.54321106229346094451671)
(4,1.75324524592178132564868)
(5,1.97842026006550320005317)
(6,2.20548889843535711463574)
(7,2.43410317434556677795854)
(8,2.67962230739822082625778)
(9,2.91865313406015935981941)
(10,3.16230034422007122429932)
(11,3.39133728237966858287677)
(12,3.57922814629814034503852)
(13,3.73460622518664653313181)
(14,3.85039509462397377959064)
(15,3.93210488535635516171851)
(16,3.97116485444158939799308)
(17,3.98990736007470037094436)
(18,3.99760388338101035898831)
(19,3.99958677301795439973375)
(20,3.99995861063385804712311)
(21,3.99999996078090669371363)
(22,3.99999999999501421044101)
(23,3.99999999999999911182158)
(24,4.00000000000000000000000)
(25,4.00000000000000000000000)
(26,4.00000000000000000000000)
(27,4.00000000000000000000000)
(28,4.00000000000000000000000)
(29,4.00000000000000000000000)
    };
    \addlegendentry{\footnotesize{Shannon limit for 16-QAM}};
    \end{axis}
    \end{tikzpicture}
    \caption{}
\end{subfigure}
        \caption[ ]
        {MLP equalizer based on the regression (MSE) for a pure AWGN channel with a 16QAM constellation. (a) represents the AIR and MSE metrics for a validation dataset over the training phase; (b) shows the AIR of a testing dataset when using the trained MLP model in (a) when using the weights of the epoch that produced maximum validation AIR (AIR early stopping) or when using the weights that produced minimum validation MSE (standard MSE minimization). } 
        \label{fig:study_AWGN_laurent}
\end{figure*}
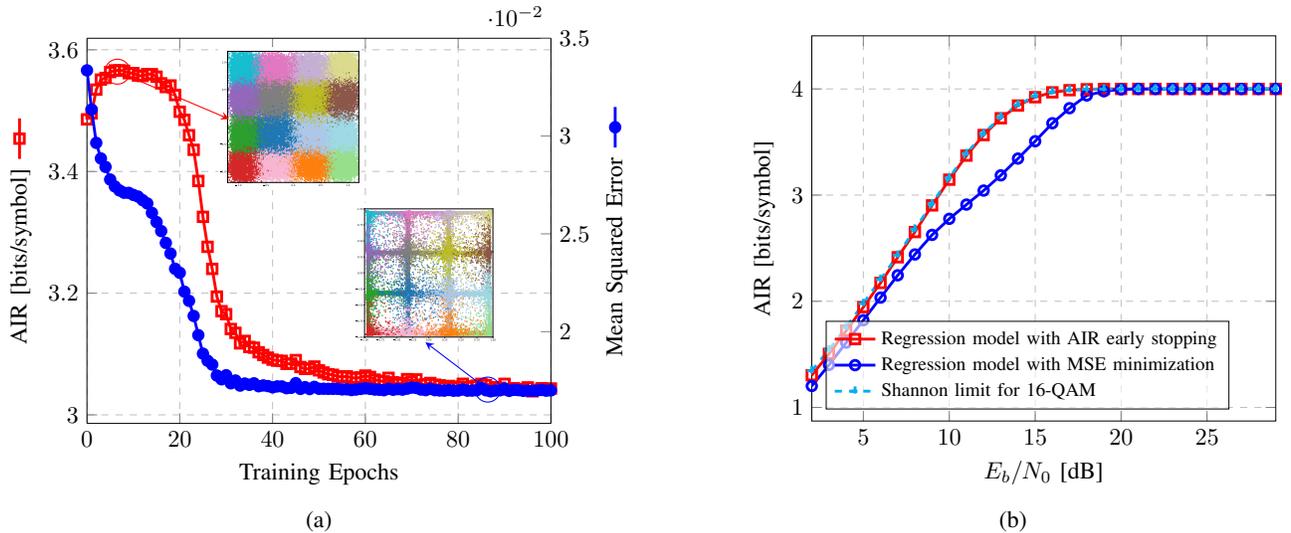

\subsection{Regression-based Equalizers and Multi-Class Classifiers}

Equalization is the task of recovering the transmitted data $X_n$ from the received data $Y_n$. It maps $Y_n$ to the most likely transmitted data, $\hat{X}_n=f(Y_n;\Theta)$, for a given mapping function $f(\cdot)$ and the set of trainable parameters $\Theta$, optimized according to some likelihood measure. In our case of NN-based equalization, 
$f(\cdot;\Theta)$ denotes the NN itself, and $\Theta$ denotes its trainable weights and biases.
Here, $X_n$ and $Y_n$ can denote either a single sample of transmitted and received data or the sequences of samples for either one or both of them (see the explicit explanations of our NN structure below in Subsec.~\ref{subsec:NN}). For simplicity of presentation, we assume the single sample representation, and that $X_n$ is chosen from a constellation alphabet $\{c_1,c_2,...,c_m\}$ with $c_i\in\mathbb{C}$, the complex space. 

In regression-based equalization, $f(Y_n;\Theta)$ is relaxed to $f_{\rm reg}(Y_n;\Theta)$ outputting any complex value, and its likelihood maximization boils down to the minimization of MSE, i.e. to finding the specific set of parameters $\Theta^*_{\rm reg}$:
\begin{equation}\label{eq:reg}
\Theta^*_{\rm reg} = \text{argmin}_{\Theta}\left\{\mathbb{E}_{X_n,Y_n}\left[|X_n-f_{\rm reg}(Y_n;\Theta)|^2\right]\right\}.
\end{equation}
The above expectation $\mathbb{E}_{X_n, Y_n}$ is taken over the samples of transmitted data and the corresponding received data. Those samples are, in fact, distributed according to $P(X_n,Y_n)=P(X_n)P(Y_n|X_n)$, where $P(X_n)$ is the transmitted signal distribution and $P(Y_n|X_n)$ describes how likely the channel output $Y_n$ is upon the transmission of $X_n$. 

In the classification-based equalization, i.e. effectively combining the regression with soft demapping into a single NN, we have: $\hat{X}_n\in\{c_1,c_2,...,c_m\}$.
In this case, $f(Y_n;\Theta)$ is relaxed to $f_{\rm cl}(Y_n;\Theta)$ outputting a vector of posterior probabilities $(q_1,...,q_m)$, where $q_k:=Q(X_n=c_k|Y_n;\Theta)$, showing how likely $X_n=c_k$ are, given receiving $Y_n$. Then, $\hat{X}_n$ is equal to the $c_k$ that has the largest posterior probability. It turns out that the ``maximum likelihood'' estimation (the best effort of the model $f_{\rm cl}(\cdot)$) is obtained if the following categorical CEL of the actual posterior $P(X_n|Y_n)$ and $Q(X_n|Y_n;\Theta)$ is minimized:
\begin{equation}\label{eq:CCELfunction}
    \mathcal{X}(P,Q;\Theta)=-\mathbb{E}_{Y_n}\left[\sum_{k=1}^m P(c_k|Y_n)\log_2(Q(c_k|Y_n;\Theta))\right].
\end{equation}

Equivalently, one can instead maximize
\begin{equation}\label{eq:mi_cl}
I_\Theta(X_n;\hat{X}_n) =  \mathbb{E}_X[\log_2(P(X_n))]- \mathcal{X}(P,Q;\Theta),
\end{equation}
where $I_\Theta(X_n;\hat{X}_n)$ is the achievable information rate for the mismatched decoding rule $Q(X_n|Y_n;\Theta)$ \cite[Def. 12]{sadeghi2009optimization},\cite[Theorem 2]{bocherer2019probabilistic}. As a result, the classification-based equalization is optimized for the following set of parameters:
\begin{equation}\label{eq:clas}
    \Theta^*_{\rm cl} = \text{argmax}_\Theta\left\{I_\Theta(X_n;\hat{X}_n)\right\}.
\end{equation}
Note that $I_\Theta(X_n;\hat{X}_n)\leq I(X;Y)$, the true mutual information of the channel, and the equality holds if $Q(X_n|Y_n;\Theta^*_{\rm cl})=P(X_n|Y_n)$ for all $(X_n,Y_n)$.
 
 In the case of regression-based equalization, the AIR cannot be expressed directly from the optimization cost, Eq.~\eqref{eq:reg}. Instead, we computed an AIR using the Gaussian approximation of conditional probabilities (a mismatched distribution) by
 \begin{equation*}
  \tilde{Q}(\hat{X}_n |X_n=c_k) = \frac{1}{2\pi\sqrt{\det(\Sigma_k)}}
  \exp\left(-\frac{1}{2} Z_{n,k}\Sigma_k^{-1}Z_{n,k}^T   \right),
 \end{equation*}
 where $Z_{n,k}=(\text{Re}\{\hat{X}_n-\mu_k\},\text{Im}\{\hat{X}_n-\mu_k\})$ and 
 $\Sigma_k=\mathbb{E}[Z_{n,k}^T Z_{n,k}|X_n=c_k]$, with trainable mean $\mu_k$ and covariance matrix $\Sigma_k$. Correspondingly, we define
 \begin{equation} \label{eq:MI-approx}
     \tilde{I}_{\rm reg}(X_n;\hat{X}_n)=\mathbb{E}_{X_n,Y_n}\left[\log_2\left(
     \frac{\tilde{Q}(\hat{X}_n |X_n)}{\sum_{k=1}^m P(c_k)\tilde{Q}(\hat{X}_n |c_k)}\right)\right].
 \end{equation}
 We use this achievable information rate to compare with the classification-based equalizer in \eqref{eq:mi_cl}.
However, one should take into account that $\tilde{I}_{\rm reg}(X_n;\hat{X}_n)$ \textit{underestimates the true mutual information} $I(X_n;\hat{X}_n)$~\cite{sadeghi2009optimization,merhav1994information}.
 
\begin{figure*}[ht!]    
\begin{subfigure}[t]{0.5\textwidth}
            \centering
                \begin{tikzpicture}[scale=0.9]
      \begin{axis} [
        xlabel={$E_b/N_0$ [dB]},
        ylabel={BER},
        grid=major,   
    	xmin=3, xmax=15,
    	xtick={4,6,8,10,12,14},
      ymode=log,
        legend style={legend pos=south west, legend cell align=left,fill=white, fill opacity=0.6, draw opacity=1,text opacity=1},
    	grid style={dashed}]
        ]
           \addplot[color=red, mark=square, very thick]
    coordinates {
     (4,  0.11904462178548177) (6,  0.08419736226399739) (8, 0.05311775207519531) (10,  0.027033487955729168) (12,  0.010126749674479166) (14, 0.0024127960205078125)
    };
    \addlegendentry{\footnotesize{MLP - CEL}};

    \addplot[color=blue, mark=triangle, very thick]        coordinates {
     (4, 0.12016487121582031) (6, 0.08479817708333333) (8,  0.05318005879720052) (10, 0.027421951293945312) (12, 0.010189056396484375) (14, 0.0022862752278645835)
    };
    \addlegendentry{\footnotesize{MLP - MSE}};
    
               \addplot[color=green, mark=o, very thick]
    coordinates {
     (4, 0.13014094034830728) (6, 0.099853515625) (8, 0.0740801493326823) (10, 0.05285835266113281) (12, 0.0371246337890625) (14, 0.025569915771484375)
    };
    \addlegendentry{\footnotesize{AWGN + Dispersion}};

    \addplot[color=cyan, mark=+,dashed, very thick]        coordinates {
     (4, 0.11877690292734486) (6, 0.08405329779466206) (8,  0.052579744999821286) (10, 0.02661251614813907) (12,  0.009787814729145285) (14, 0.002223720020628769)
    };
    \addlegendentry{\footnotesize{Ideal AWGN}};
    \end{axis}
    \node at (1.5,3) {\includegraphics[width=2.5cm]{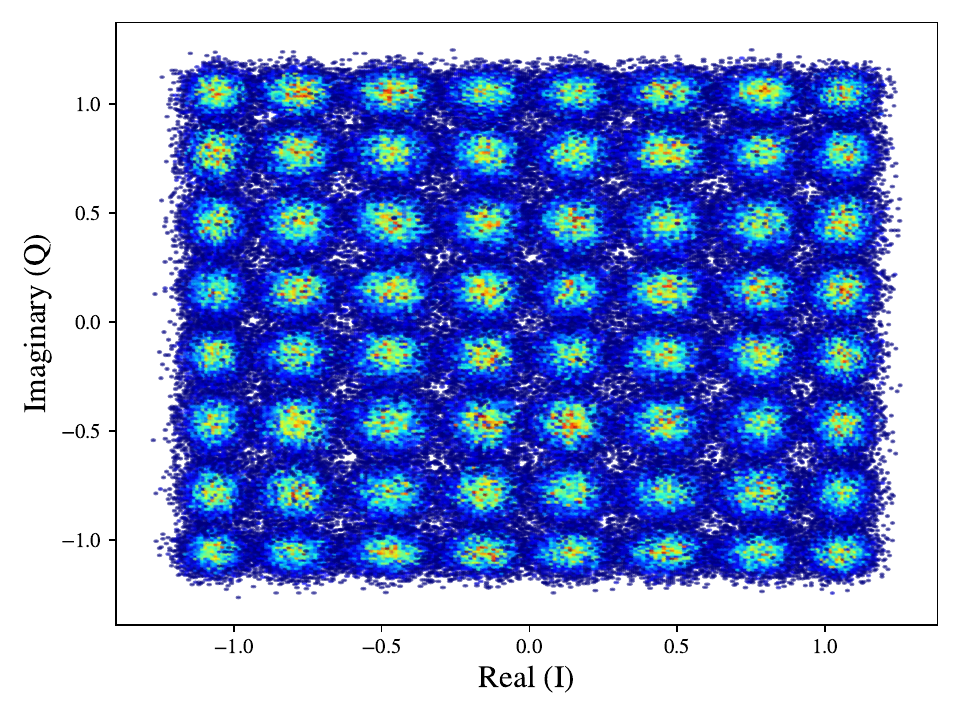}};
\node (mark) [draw, red, circle, minimum size = 5pt] at (5.12, 2.2){};
    \draw [-stealth,red](5.12, 2.2) -- (2.9, 3);

    \end{tikzpicture}
    \caption{}
\end{subfigure}
\hspace{\fill}
\begin{subfigure}[t]{0.5\textwidth}
            \centering
                \begin{tikzpicture}[scale=0.9]
      \begin{axis} [
        xlabel={$E_b/N_0$ [dB]},
        ylabel={AIR [bits/symbol]},
        grid=both,   
    	xmin=3, xmax=15,
    	xtick={4,6,8,10,12,14},
        legend style={legend pos=south west, legend cell align=left,fill=white, fill opacity=0.6, draw opacity=1,text opacity=1},
    	grid style={dashed}]
        ]
               \addplot[color=red, mark=square, very thick]
    coordinates {
     (4, 3.748626884332882) (6,  4.319834066522084) (8,  4.882362144235495) (10, 5.397181460677634) (12, 5.762630534179394) (14, 5.940344361771844)
    };
    \addlegendentry{\footnotesize{MLP - CEL}};

    \addplot[color=blue, mark=triangle, very thick]        coordinates {
     (4, 3.7400359906476717) (6, 4.320410803368097) (8, 4.871949107460307) (10, 5.3937195171203145) (12, 5.770759276558682) (14, 5.9415758281652264)
    };
    \addlegendentry{\footnotesize{MLP - MSE}};
    
               \addplot[color=green, mark=o, very thick]
    coordinates {
     (4,  3.5921402241304947) (6, 4.069758344960417) (8,  4.492478035746393) (10, 4.916678754543221) (12,  5.210905508944452) (14, 5.444515546995937)
    };
    \addlegendentry{\footnotesize{AWGN + Dispersion}};

    \addplot[color=cyan, mark=+,dashed, very thick]        coordinates {
     (4, 3.76989845]) (6, 4.34379197]) (8, 4.90643037) (10, 5.40385792) (12, 5.77360049) (14, 5.94012531)
    };
    \addlegendentry{\footnotesize{Ideal AWGN}};
    \end{axis}
    \end{tikzpicture}
    \caption{}
\end{subfigure}
        \caption[ ]
        {MLP NN based on the regression (MSE) and multi-class classifier (CCEL) investigation over the BER (a) and AIR (b) metrics for the AWGN system with 3 km linear fiber dispersion with 64 QAM. The constellation for validation dataset after the MLP-MSE equalizer with AIR early stopping is presented for the toy case when $E_b/N_0 = 12$ dB. } 
        \label{fig:study_AWGN}
\end{figure*}
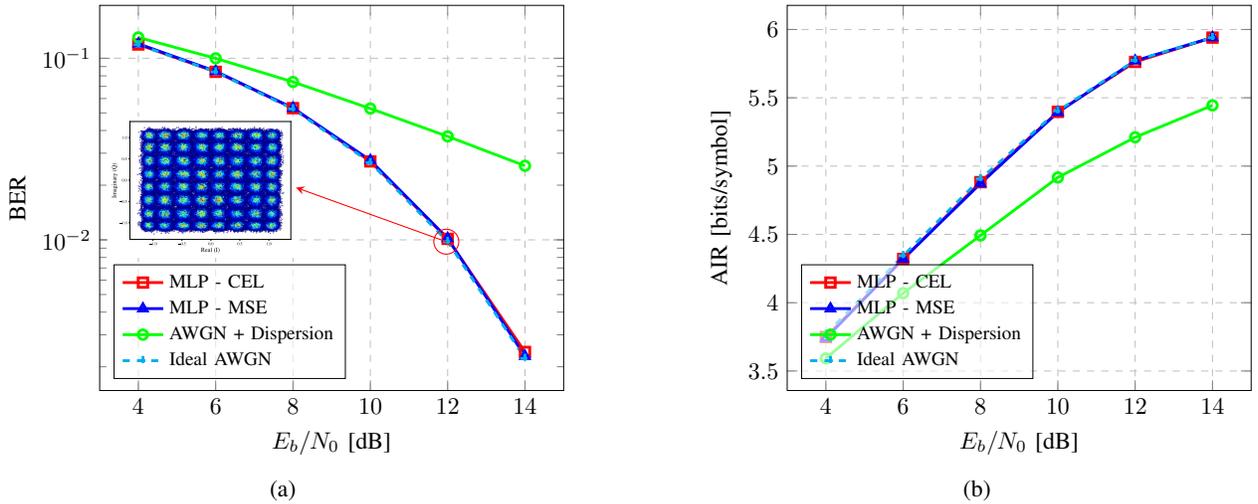

 Note that it is known that the regression-based equalization in \eqref{eq:reg} can be expressed as a classification-based one in \eqref{eq:clas} with some special $f_{\rm cl}(Y_n;\Theta)$ outputting posterior probabilities based on a complex Gaussian distribution:
\begin{equation}\label{eq:Gaus}
  Q_{\rm G}(X_n|Y_n)  = \frac{1}{2\pi \sigma^2}
  \exp{\left(-\frac{|X_n-f_{\rm reg}(Y_n;\Theta)|^2}{2\sigma^2}\right)},
\end{equation}
with some variance of $\sigma^2$ for each real and imaginary tributary. Accordingly,
\begin{equation}
    \log_2(Q_{\rm G}(X_n|Y_n))=-\frac{|X_n-f_{\rm reg}(Y_n;\Theta)|^2}{2\ln(2)\sigma^2}-\log_2(2\pi\sigma^2),
\end{equation}
and 
\begin{align}
    \mathcal{X}(P,Q_{\rm G};\Theta)&= \frac{1}{2\ln(2)\sigma^2}
    \mathbb{E}_{X_n,Y_n}\left[|X_n-f_{\rm reg}(Y_n;\Theta)|^2\right]\nonumber\\
    &\hspace{3cm}+\log_2(2\pi\sigma^2).
\end{align}

One can verify that $\mathcal{X}(P,Q_{\rm G};\Theta)$ is minimized if $\Theta=\Theta^*_{\rm reg}$ and we choose $\sigma_{\rm reg}^2=\frac{1}{2}
\mathbb{E}_{X_n,Y_n}\left[|X_n-f_{\rm reg}(Y_n;\Theta^*_{\rm reg})|^2\right]$, the MSE.
However, we need to point out one important pitfall already reported in Ref.\cite{freire2022neural}. Regarding the quality of transmission metrics, when performing the regression task via the MSE, the equalization model, if not monitored, starts to distort the QAM constellation into a so-called ``jail window'' pattern. The consequence of this is that the AIR calculation of such a new constellation becomes underestimated if using a Gaussian demapper  with $\sigma_{\rm reg}^2$. At this point, two things can be done to solve this problem: i) we can use an optimizer (e.g. linear search algorithm) to find a new $\sigma^2$ that maximizes the achievable information rate or even embed in the original NN the demapping process to learn this step jointly using the CEL loss function; ii) we can monitor the AIR of the model for a validation dataset and apply early stopping considering not the MSE loss function, \textit{but the AIR metric itself}\footnote{Another alternative, recently proposed in Ref.~\cite{diedolo2022nonlinear}, can also be used to solve the ``jail window'' problem: it suggests incorporating an entropy regularization into the MSE function. This loss  is referred to as the MSE-X loss function.}. Solution ii) is simpler and produces an NN architecture that is less computationally complex compared to solution i).

To illustrate the effectiveness of solution ii) and the impact of the ``jail window'' constellation, we trained the MLP equalizer on a pure AWGN channel. The codes for this test with specific details are available in Ref.~\cite{pedro_jorge_freire_2022_6240318}. Fig.~\ref{fig:study_AWGN_laurent} summarizes the analyses done for this first investigation. In Fig.~\ref{fig:study_AWGN_laurent}~(a), the MLP model is trained with a bit-energy to noise power ratio ($E_b/N_0$) equal to 12~dB, and a validation dataset is used to monitor both the AIR and MSE over 100 epochs. As can be seen, initially the AIR increases and the MSE decreases until a certain epoch when the AIR and MSE become almost constant. However, when we continue to train the NN, the MLP equalizer finds a set of weights that can further decrease the MSE, but this does not result in any additional gain in the AIR, but the AIR tendency is quite opposite because now the solution produces the ``jail window'' constellation. We have highlighted in Fig.~\ref{fig:study_AWGN_laurent}~(a) the constellations when the model is saved on the highest validation AIR, and when the model is saved on the lowest validation MSE. To show the consequence of such a wrong early stopping in the testing performance, we have tested the model that we trained at $E_b/N_0=12$~dB, with $E_b/N_0$ varying from 2~dB to 29~dB, where we used the weights for the AIR early stopping and the ``pure'' MSE minimization. Fig.~\ref{fig:study_AWGN_laurent}~(b) shows the result of our scrutiny, where we can see that using the model with the MSE minimization and $\sigma_{\rm reg}^2$ for demapping produced the worst AIR value in the testing dataset than the situation when we used the model with the AIR early stopping and the same $\sigma_{\rm reg}^2$ for the demapping. Option ii) is the method utilised by the regression models in this paper, and as seen in the Fig.~\ref{fig:study_AWGN_laurent} (b), it reached the same AIR as the Shannon limit for 16 QAM.

To show further the empirical confirmation that when using the AIR early stooping, the performance of the regression model does not get reduced compared to option i), where we have a NN model that learns the demapping jointly, we conduct the following investigation. We have simulated the simple 32 GBd, 64QAM transmission, where 3~km of linear fiber with the second-order dispersion parameter $D = 17$ ps/(nm$\cdot$km) is included to add some small amount of linear dispersion (losses and nonlinearities are not considered) and after the fiber, an AWGN is added with $E_b/N_0$ varying from 4~dB to 14~dB. At the receiver, we consider a pulse-shaping filter, and we have the same NN-based MLP as described in Section II-B, operating with 1 sample/symbol, to recover the dispersion introduced by the fiber. With this experiment, we want to show that the NN equalization based on the MSE with AIR early stopping and the NN based on the CEL will provide the same AIR and BER as we have for the system with only the AWGN and no fiber (we mark the latter as ``Ideal AWGN''). 

In Fig.~\ref{fig:study_AWGN}, the best values of AIR and BER achieved by the NN equalizers/soft-demapping based on either the MSE or the CEL are presented for different levels of $E_b/N_0$, when using the same NN training procedure described in Sec II.D. We also plotted the equalized constellation histograms after the NN based on the MSE for an exemplary point $E_b/N_0$=12~dB using a testing dataset. This constellation shows that the symbols after the NN based on MSE retain the form of the AWGN-induced constellation points histogram. As expected, both loss functions led to learning the linear chromatic dispersion filter quite precisely, and both NNs reached the performance of the ``Ideal AWGN'' system. This empirical example shows that a simple linear problem governed by AWGN can be learned using either the MSE with AIR early stopping or the CEL.  

At the same time, the situation in which we consider a long nonlinear fiber transmission differs from the above simple case.
Since we use an optimization algorithm based on gradient descent, the training can also \textit{suffer from many NN-learning-related pitfalls}. Therefore, the goal of our paper is to show that this statistical limitation/mismatch of the MSE in regression can be ``less harmful'' to the AIR (or any eventual quality metric used) than the gradient learning-related problems that the CEL can (and typically does) produce in the case of classification. In short, our main result is that the NN learning/training-related problems typically downgrade the ultimate quality metrics more than the MSE distribution-related mismatch, insofar as it is virtually impossible to achieve the ``ideal'' training in the CEL classification case in reality, and the resulting penalty ``always wins'' (again, in contrast to the simple 3 km linear fiber case presented above).

\begin{figure}[ht!]
   \centering
    \includegraphics[width=0.4\textwidth]{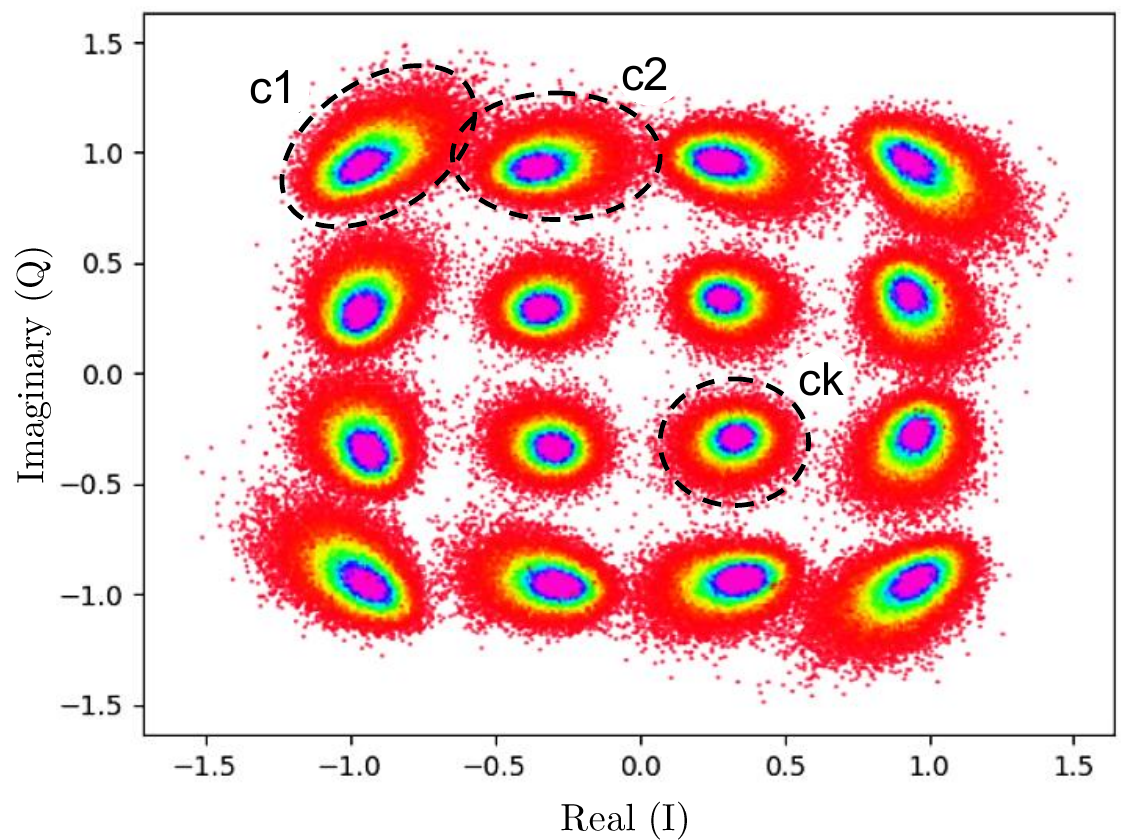}
     \caption{Distribution of the received symbols in an ideal transmitter of SC-DP 16QAM case with launch power -1 dBm and 34.4 GBd, over 9$\times$50 km TWC fiber. The constellation is given after a standard Rx-DSP described in Sec \ref{sec:simul}.}
    \label{fig:constellation}
\end{figure}

To make some general notes, as mentioned previously, when AWGN is the primary source of signal distortion the regression-based linear equalization with MSE loss, Eq.~\eqref{eq:reg}, can be quite an appropriate choice in terms of training \cite{Georg2021book}. However, such equalizers are penalized if the distortion statistics deviate from the AWGN statistics. This typically occurs in optical communication, as the optical fiber is nonlinear and dispersive. An example of the effect of nonlinear distortions is shown in the received constellation in Fig.~\ref{fig:constellation}, where the transmission of 16-QAM data is simulated over a particular optical link. The simulation and transmission setup are detailed later in Sec.~\ref{sec:simul}. We observe here that because of nonlinear effects, the distortion of each constellation cluster has visually different statistics. 

On the one hand, the classification-based equalizer
could potentially outperform a regression-based equalizer (in a proper scenario) as it can adapt itself to any channel statistics, so, compared to the MSE regression, the classifier's functioning does not depend on how close the resulting distribution is to the Gaussian one. In particular, $I_{\Theta^*_{\rm cl}}(X_n;\hat{X}_n)$ can approach the true $I(X;Y)$ provided that the classification equalizer $f_{\rm cl}(\cdot;\Theta)$ has large enough ``learning complexity/capacity'' (e.g. number of parameters and flexibility of $f_{\rm cl}(\cdot;\Theta)$ are sufficient to represent the given phenomenon). However, a classification-based equalizer can also have drawbacks. For instance, one potential drawback is that the CEL function is independent of the spatial proximity of constellation points involved in the decision (classification) process. In other words, the CEL penalizes the misclassification between two classes with the same  ''cost''  value with disregard to the ``type'' of misclassification occurrence, and this can degrade the NN learning process. This issue is discussed in more detail in Ref.~\cite{lathuiliere2019comprehensive}.

Yet another disadvantage of CEL minimization is that according to Ref.~\cite{bosman2020visualising}, the CEL surfaces have fewer local minima than the square error-based losses (SEL), to which the MSE loss belongs. However, the CEL landscape is typically prone to sharp local minima, while for the MSE, these are usually the wide local minima. Such sharp local minima produced by the CEL, exhibit stronger gradients in low training error regions than the MSE does, which causes the infamous overfitting in the systems trained with the CEL, resulting in MSE having a better generalization property in almost all the cases tested in \cite{bosman2020visualising}.\footnote{
Another drawback of classification-based systems is that it is tailored to each task, i.e. it has to have a specified fixed number of outputs corresponding to the constellation's cardinality. This indicates that the classifier model's operation is specific to the modulation format on which it was trained, the feature that inhibits the practical (say, hardware) classifier implementation. But in the case of regression~\cite{freire2021transfer,pedroTF}, we do not need to retrain the model at all for it to work on other modulation formats. Because of this, the regression model is far more adaptive than the classification one.} In the next section, we compare the performance and the training pitfalls related to the depth of minima for the classification and regression in the framework of soft-demapping in optical communications. 

\subsection{Deep Neural Network Design}\label{subsec:NN}

\begin{figure}[ht!]
    \centering
    \includegraphics[width=0.5\textwidth]{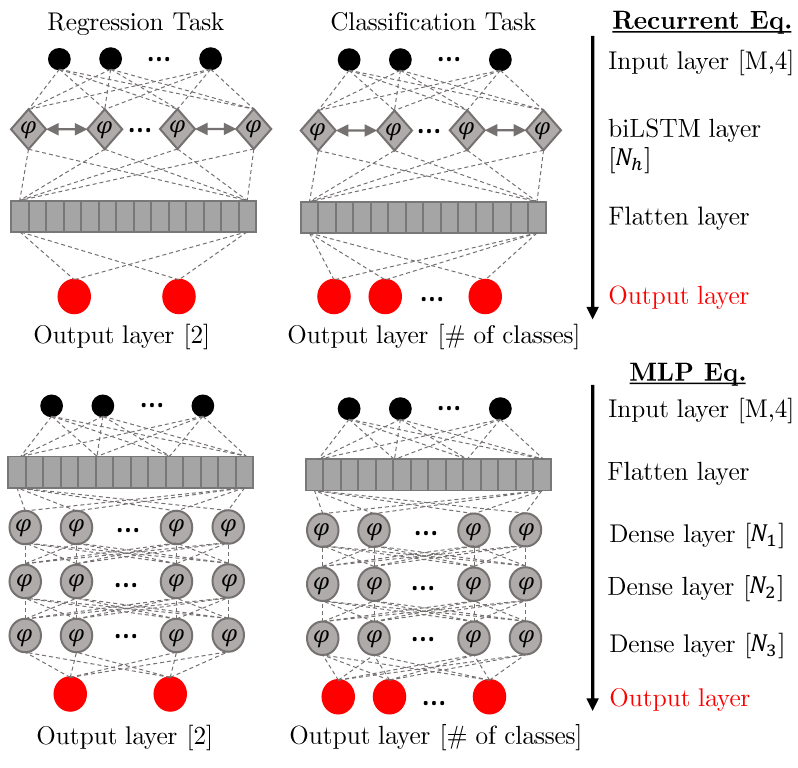}
    \caption{The schematics of different NN architectures considered in our paper. At the top, we show the regression and classification systems based on the recurrent equalizer with $N_h$ hidden units. At the bottom, we show both tasks implemented with the MLP equalizer having three hidden layers, with $N_1$, $N_2$, and $N_3$ neurons in each consecutive layer, respectively. In all cases, the output is marked with red to highlight the difference in the regression- and multi-class classifier-based approaches. For our case, the activation function $\varphi$ is ``tanh''. The remarks at the right depict the details of each structure.}
    \label{fig:NN_tasks}
\end{figure}

Before moving on to the results section, we discuss how we make the comparison of regression and classification as fair as possible.  First, we use two types of equalizers: the first is based on the feed-forward multi-layer perceptron (MLP) with three hidden layers, while the second one is the recurrent structure, consisting of one layer of bidirectional Long short-term memory (biLSTM). The comparison of these equalizers' complexity and functioning is given in Ref.~\cite{freire2021performance}. These two cases are taken to demonstrate that our outcomes are true for different NN architectures, but we note that the biLSTM layer has demonstrated better performance in previous studies \cite{freire2021performance}. The only differences between using each architecture for regression or classification tasks occur in i) the structure of the output layer, as shown in Fig~\ref{fig:NN_tasks}, and ii) in the loss function type used for each task. In the case of regression, the output layer has two linear neurons referring to the real and imaginary parts of the recovered symbol, and the loss function used is the MSE. For the multi-class classification, the number of neurons in the output layer is determined by the modulation format cardinality ($\text{MF}$ or $\#\text{QAM}$), and the NN structure ends with the softmax layer, while the loss function is the categorical CEL. We point out once again that besides these two differences, the regression and classifier models that we compare, share the same number of inputs, hidden layers, neurons in each layer, and hyperparameter values; the training/test datasets are also the same. As for the memory sizes, for both types of prediction modeling, we used the same memory length: $M=51$ for the SSMF case, and $M=41$ for the TWC case. The values of the NN parameters used in this study are given in Table.~\ref{Table_design}. Furthermore, in terms of training and testing datasets, both NN models were trained and tested with the same datasets, where for training the original training dataset had $2^{20}$ input points, and at every epoch, we picked $2^{18}$ random input points (out of total $2^{20}$) to train the NN efficiently\footnote{We observed that this training methodology improves the generalization of the training and this was also observed by similar works \cite{hoffer2020augment,KumawatKumawat,ke2018lirs}}.  For testing, we used a never-seen dataset with $2^{18}$ input points.

As we are dealing with different loss functions, we must consider that the learning rate and the number of epochs required, may differ for the regression compared to the classification. To address this potential issue, we optimize the learning rate using values in the range [$10^{-3}$, $5 \cdot 10^{-4}$, $10^{-4}$, $5 \cdot 10^{-5}$], and use the early stop to obtain the architecture that performed best during the training process. In general, the early stop was used if no improvement was seen after 150 epochs out of the total 5000 training epochs.

Additionally, in this work, all NN-based equalizers were implemented in TensorFlow (2.2.0) GPU backend and Keras (2.3.1). Moreover, the Flatten layer, being a part of a Keras API, is used to interconnect the multidimensional (3D) layers with linear (2D - Dense) layers inside the NNs architectures. The codes and dataset samples can be downloaded in Ref.~\cite{pedro_jorge_freire_2022_6240318}.

\begin{table}[htbp] 
  \centering
  \caption{ The common regression and classification tasks NN/training/testing (hyper)parameters used in our study.}
\begin{tabular}{|c|c|c|c|c|}
\hline
\textbf{Equalizer} & \textbf{Mini-Batch} & \textbf{$N_h$} & \textbf{$N_1$ / $N_2 $ / $N_3$} & \textbf{\begin{tabular}[c]{@{}c@{}}Training / Testing \\ Dataset size\end{tabular}} \\ \hline\hline
Recurrent          & $4331$              & $226$           & -                           & $10^{18}$ / $10^{18}$                    \\ \hline
MLP                & $4331$              & -              & $481$ / $31$ / $263$             & $10^{18}$ / $10^{18}$                    \\ \hline
\end{tabular}
\label{Table_design}
\end{table}

\section{Simulating Signal Propagation in Coherent Optical Transmission Systems}\label{sec:simul}

To illustrate the effects addressed in our work, we numerically simulated the dual-polarization (DP) transmission of a single-channel signal propagation at a 34.4 GBd rate. First, a bit sequence was generated using the Mersenne twister generator \cite{matsumoto1998mersenne}, which has a periodicity equal to $2^{19937} - 1$. Then, the signal is pre-shaped with a root-raised cosine (RRC) filter with 0.1 roll-off at an upsampling rate of 8 samples per symbol. In addition, the signal could have three possible modulation formats: 16 / 32 / 64-QAM. To cover different physical scenarios, we consider the following two test cases: (i) the transmission over the optical link consisting of $9\!\times\!50$~km true-wave classic (TWC) spans; and (ii) the transmission over $5\!\times\!100$~km of standard single-mode fiber (SSMF) spans. The optical signal propagation along the fiber was simulated by solving the Manakov equation via split-step Fourier method ~\cite{agrawal2013nonlinear} with the resolution of $1$ km per step\footnote{The ultra-fine step resolution guarantees that we truly model the optical channel properties captured by the NN and do not address some by-side simulation effects.}. The parameters of the TWC fiber are: the attenuation parameter $\alpha = 0.23$ dB/km, the dispersion coefficient $D = 2.8$ ps/(nm$\cdot$km), and the effective nonlinearity coefficient $\gamma = 2.5$~(W$\cdot$km)$^{-1}$. The SSMF parameters are: $\alpha = 0.2$ dB/km, $D = 17$ ps/(nm$\cdot$km), and $\gamma = 1.2$ (W$\cdot$km)$^{-1}$. The purpose of testing two different fibers is to see if the MSE-based regression task works better in the SSMF transmission because, due to the higher dispersion and lower nonlinearity, in that case, the constellation point distributions should be closer to Gaussian; for the TWC we have $6$ times lower dispersion and $2$ times higher nonlinearity, such that the non-Gaussian constellations should become more pronounced.

\begin{figure}[ht!]
    \centering
    \includegraphics[width=0.5\textwidth]{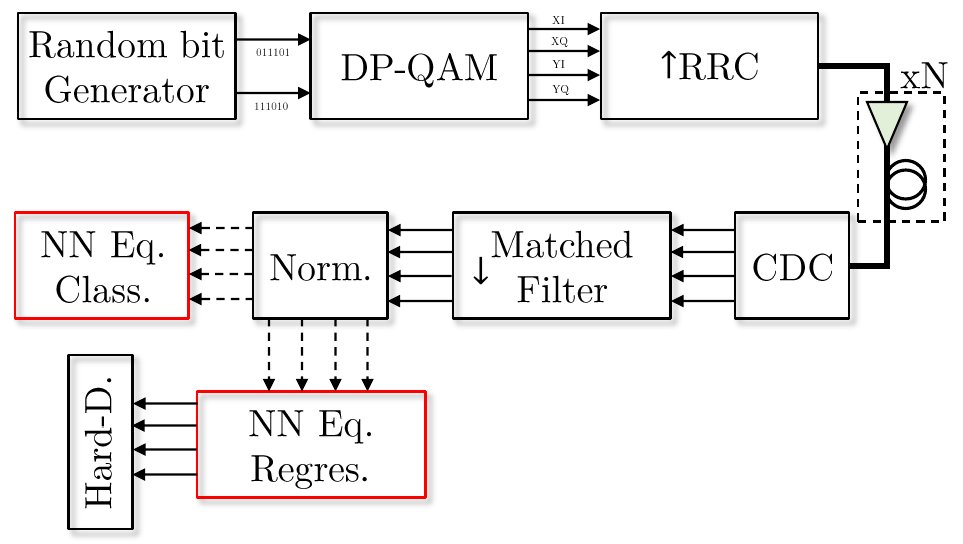}
    \caption{The schematic of the setup used in our simulations. The two available equalization/soft-demapping types (multi class classification and regression) are inserted at the receiver side after the matching filter and the DSP blocks: CDC and Phase/Amplitude Normalization. The equalizers are highlighted with a red box.}
    \label{fig:setup}
\end{figure}

In our model, every span is followed by an optical amplifier with the noise figure $\text{NF}=4.5$~dB, which fully compensates for the fiber losses and adds the ASE noise. At the receiver, a standard Rx-DSP was used. It includes the full electronic chromatic dispersion compensation (CDC) using a frequency-domain equalizer, the application of a matched filter, and downsampling to the symbol rate. Finally, the received symbols were normalized (by phase and amplitude) to the transmitted ones, i.e. multiplied by a constant $\mathcal{K}_\text{DSP} \in \mathbb{C}$ minimizing the mean squared error between the transmitted $X_n$ and the received $Y_n$ signals:
\begin{equation}
    \label{eqn:norming}
    \mathcal{K}_\text{DSP} = \min_\mathcal{K}\left\|\mathcal{K}\cdot Y_n - X_n\right\|.
\end{equation} 
After Rx-DSP, the output symbols were processed by an NN-based equalizer for further signal enhancement. Fig.~\ref{fig:setup} shows all the blocks involved in the transmission simulations, where we highlight regression/classification-based NN equalizers/soft-demapping with red boxes. In addition to the AIR, another performance metric used in this paper is the Q-factor calculated directly from the BER values after the hard decision as:
\begin{equation}\label{eq:eq.Vadd}
\begin{split}
Q = 20 \log_{10}\left[\sqrt{2}~\textrm{erfc}^{-1}(2* \text{BER})\right],
\end{split}
\end{equation}
where $\text{erfc}^{-1}$ is the inverse complementary error function. Note that the hard-decision block is optional: it is used for the Q-factor computation, but redundant when we deal with the AIR.

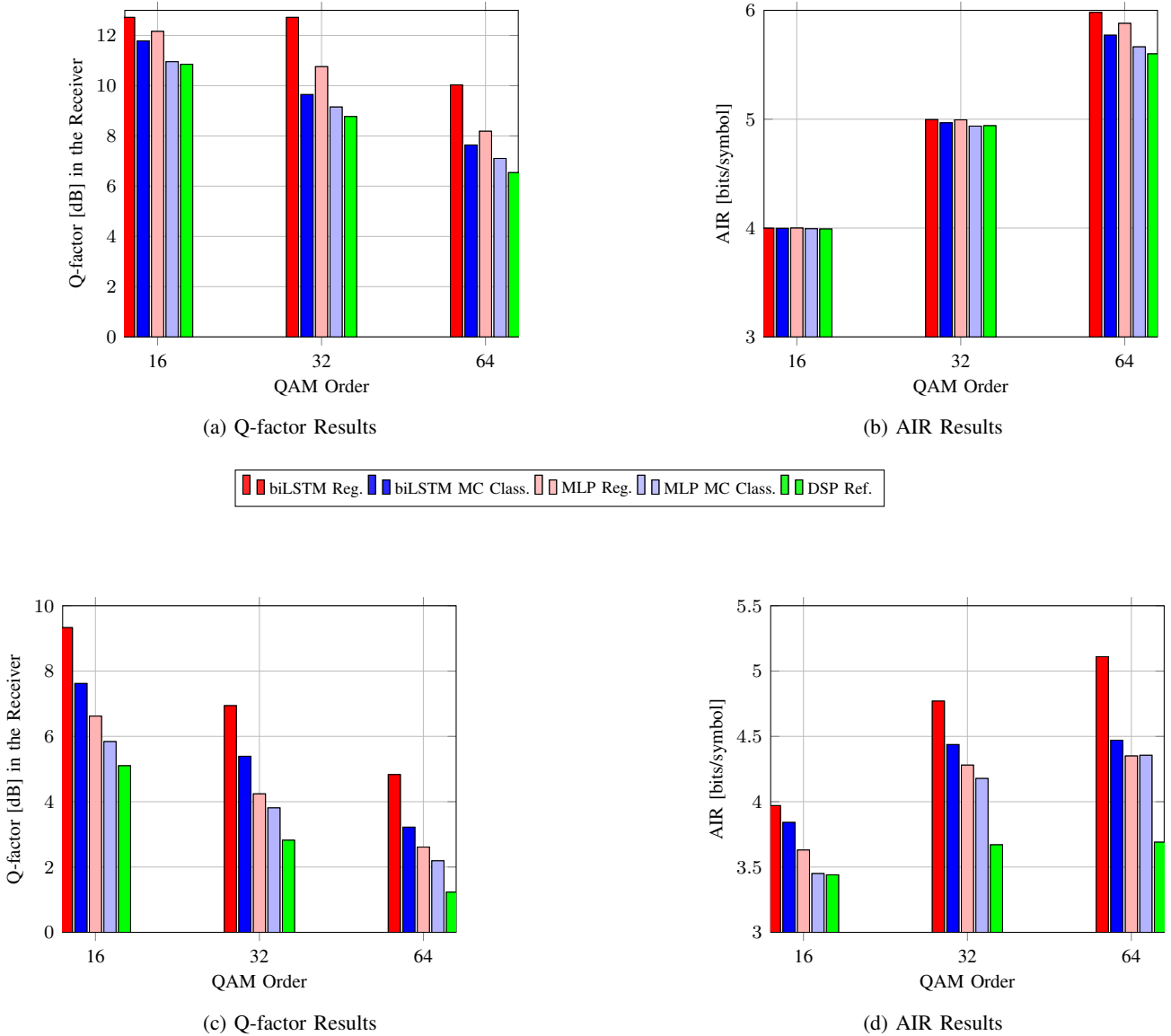
\begin{figure*}[t!]
  \centering
   \begin{subfigure}[b]{0.475\textwidth}
            \centering
                    \begin{tikzpicture}[scale=0.85]
      \begin{axis}[
        xlabel={QAM Order},
        ylabel={Q-factor [dB] in the Receiver},
        set layers,
        ybar=1.2pt,
        bar width=6pt,
        symbolic x coords={16,32,64},
        grid=both,
        ymin=0,
        samples=4,
        legend cell align=left,
        ymax=13,
        x label style={font=\footnotesize},
        y label style={font=\footnotesize},
        ticklabel style={font=\footnotesize},
        xtick={16,32,64},
        ]

                \addplot[black,fill=red,postaction=] coordinates {
       (16, 12.719) (32, 12.719) (64, 10.031) 
        };
        \addplot[black,fill=blue,postaction=] coordinates {
      (16, 11.784) (32, 9.6466) (64, 7.6396) 
        };

        \addplot[black,fill=red!30!white,postaction=] coordinates {
      (16, 12.16) (32, 10.757) (64, 8.1857) 
        };
                \addplot[black,fill=blue!30!white,postaction=] coordinates {
        (16,10.9547) (32, 9.1506) (64, 7.1032) 
        };
        \addplot[black,fill=green,postaction=] coordinates {
         (16, 10.85) (32, 8.77) (64, 6.54) 
        };
      \end{axis}
    \end{tikzpicture}

    \caption{Q-factor Results}
        \end{subfigure}
   \hfill
   \begin{subfigure}[b]{0.475\textwidth}
            \centering
                    \begin{tikzpicture}[scale=0.85]
      \begin{axis}[
        xlabel={QAM Order},
        ylabel={AIR [bits/symbol]},
        set layers,
        ybar=1.2pt,
        bar width=6pt,
        symbolic x coords={16,32,64},
        grid=both,
        ymin=3,
        samples=4,
        legend cell align=right,
        ymax=6,
        x label style={font=\footnotesize},
        y label style={font=\footnotesize},
        ticklabel style={font=\footnotesize},
        xtick={16,32,64},
        ]

                \addplot[black,fill=red,postaction=] coordinates {
     (16, 3.998) (32, 4.996) (64, 5.981) 
        };
        \addplot[black,fill=blue,postaction=] coordinates {
       (16, 3.99711461) (32,4.966673745) (64, 5.771621375) 
        };

        \addplot[black,fill=red!30!white,postaction=] coordinates {
       (16, 3.9997) (32, 4.9938) (64, 5.88) 
        };
                \addplot[black,fill=blue!30!white,postaction=] coordinates {
       (16, 3.993219333) (32,4.936278283) (64, 5.664429133) 
        };
        \addplot[black,fill=green,postaction=] coordinates {
       (16, 3.991) (32, 4.94) (64, 5.6) 
        };
      \end{axis}
    \end{tikzpicture}

    \caption{AIR Results}
        \end{subfigure}
    \vskip\baselineskip
   \begin{subfigure}[b]{0.475\textwidth}
            \centering
                    \begin{tikzpicture}[scale=0.85]
      \begin{axis}[
        xlabel={QAM Order},
        ylabel={Q-factor [dB] in the Receiver},
        set layers,
        ybar=1.2pt,
        bar width=6pt,
        symbolic x coords={16,32,64},
        grid=both,
        ymin=0,
        samples=4,
        legend cell align=left,
        ymax=10,
        x label style={font=\footnotesize},
        y label style={font=\footnotesize},
        ticklabel style={font=\footnotesize},
        xtick={16,32,64},
               legend style={at={(0.668,.8)},anchor=west,font=\scriptsize},
                legend style={at={(0.5, 1.6)},anchor=west,font=\scriptsize,fill=white, fill opacity=0.6, draw opacity=1,text opacity=1, legend columns =-1},
        ]

                \addplot[black,fill=red,postaction=] coordinates {
       (16, 9.33) (32, 6.94) (64, 4.83) 
        };
        \addlegendentry{biLSTM Reg.}
        \addplot[black,fill=blue,postaction=] coordinates {
      (16, 7.62) (32, 5.39) (64, 3.22) 
        };
        \addlegendentry{biLSTM MC Class.}

        \addplot[black,fill=red!30!white,postaction=] coordinates {
      (16, 6.62) (32, 4.24) (64, 2.61) 
        };
        \addlegendentry{MLP Reg.}
                \addplot[black,fill=blue!30!white,postaction=] coordinates {
        (16, 5.84) (32, 3.81) (64, 2.19) 
        };
        \addlegendentry{MLP MC Class.}
        \addplot[black,fill=green,postaction=] coordinates {
         (16, 5.1) (32, 2.82) (64, 1.23) 
        };
        \addlegendentry{DSP Ref.}
      \end{axis}
    \end{tikzpicture}

    \caption{Q-factor Results}
        \end{subfigure}
   \hfill
   \begin{subfigure}[b]{0.475\textwidth}
            \centering
                    \begin{tikzpicture}[scale=0.85]
      \begin{axis}[
        xlabel={QAM Order},
        ylabel={AIR [bits/symbol]},
        set layers,
        ybar=1.2pt,
        bar width=6pt,
        symbolic x coords={16,32,64},
        grid=both,
        ymin=3,
        samples=4,
        legend cell align=left,
        ymax=5.5,
        x label style={font=\footnotesize},
        y label style={font=\footnotesize},
        ticklabel style={font=\footnotesize},
        xtick={16,32,64},
        ]

                \addplot[black,fill=red,postaction=] coordinates {
     (16, 3.97) (32, 4.77) (64, 5.11) 
        };
        \addplot[black,fill=blue,postaction=] coordinates {
       (16, 3.841303546) (32, 4.437348934) (64, 4.470743257) 
        };

        \addplot[black,fill=red!30!white,postaction=] coordinates {
       (16, 3.63) (32, 4.28) (64, 4.35) 
        };
                \addplot[black,fill=blue!30!white,postaction=] coordinates {
       (16, 3.449641132) (32, 4.177663827) (64, 4.355327653) 
        };
        \addplot[black,fill=green,postaction=] coordinates {
       (16, 3.44) (32, 3.67) (64, 3.69) 
        };
      \end{axis}
    \end{tikzpicture}

    \caption{AIR Results}
        \end{subfigure}
  \caption{Performance study for regression equalizer (Reg.) and multi-class classifier (MC Class.) in different modulation format [16-QAM,  32-QAM  and  64-QAM] transmission at (a,b) 6 dBm and (c,d) 10~dBm SC-DP, 5x100km SSMF fiber and 34.4G~Bd. }
  \label{figure:diffmod}
\end{figure*}

The NN input mini-batch shape, for both regression and classification tasks, can be defined by three dimensions \cite{freire2021performance}: $(B, M, 4)$, where $B$ is the mini-batch size, $M$ is the memory size defined through the number of neighbours $N$ as $M = 2N + 1$, and $4$ is the number of features for each symbol, referring to the real and imaginary parts of two polarization components. For the regression, the output target is to recover the real and imaginary parts of the $k$-th symbol in one of the polarization, so the shape of the NN output batch can be expressed as $(B,2)$. In the case of multi-class classification, the output will provide the vector probability of a received symbol to belong to a certain class, and so the output batch shape is equal to $(B, \text{MF})$\footnote{For the case of multi-label classification as represented in Fig.\ref{fig:NN_tasks}, we have appended one extra layer with $\log_2(\text{MF})$ neurons to represent the bit probabilities. So, for this task, the output batch shape is equal to $(B, \log_2(\text{MF}))$}. Here, we would like to draw your attention to the fact that, since the NN model comes after the traditional Rx-DSP chain, it will only take care of some residual memory and its coupling with the nonlinearity, not the full memory. Finally, we note that different random seeds were used to produce training and testing datasets, ensuring their independence and avoiding overestimation, with the cross-correlation not exceeding $0.02$.

\begin{figure*}[ht!]    
\begin{subfigure}[t]{0.5\textwidth}
            \centering
                \begin{tikzpicture}[scale=0.9]
      \begin{axis} [
        xlabel={Launch power [dBm]},
        ylabel={Q-factor [dB]},
        grid=both,   
    	xmin=-5, xmax=4,
    	xtick={-5, ..., 4},
        legend style={legend pos=south west, legend cell align=left,fill=white, fill opacity=0.6, draw opacity=1,text opacity=1},
    	grid style={dashed}]
        ]
           \addplot[color=red, mark=square, very thick]
    coordinates {
     (-5, 7.826) (-4, 8.21) (-3, 8.55) (-2, 8.91) (-1, 9.17)
     (0, 9.43) (1, 9.6) (2, 9.69) (3, 9.6) (4, 9.36)
    };
    \addlegendentry{\footnotesize{biLSTM Reg. (Test)}};

    \addplot[color=blue, mark=triangle, very thick]        coordinates {
    (-5, 7.77) (-4, 8.11) (-3, 8.28) (-2, 8.54) (-1, 8.83)
     (0, 8.86) (1, 8.76) (2, 8.89) (3, 8.88) (4, 8.6)
    };
    \addlegendentry{\footnotesize{biLSTM MC Class. (Test)}};
    
               \addplot[color=red, mark=square,dashed, very thick]
    coordinates {
     (-5, 7.86) (-4, 8.25) (-3, 8.6) (-2, 8.96) (-1, 9.3)
     (0, 9.55) (1, 9.8) (2, 10) (3, 10.2) (4, 10.32)
    };
    \addlegendentry{\footnotesize{biLSTM Reg. (Train)}};

    \addplot[color=blue, mark=triangle,dashed, very thick]        coordinates {
    (-5, 7.97) (-4, 8.37) (-3, 8.59) (-2, 9.57) (-1, 11.51)
     (0, 13) (1, 13) (2, 13) (3, 13) (4, 13)
    };
    \addlegendentry{\footnotesize{biLSTM MC Class. (Train)}};

    \addplot[color=green, very thick]         coordinates {
    (-5, 7.83) (-4, 8.11) (-3, 8.25) (-2, 8.18) (-1, 7.8)
     (0, 7.17) (1, 6.33) (2, 5.24) (3, 3.97) (4, 2.48)
    };
    \addlegendentry{\footnotesize{Regular DSP}};

    \end{axis}
    \end{tikzpicture}
    \caption{TWC link -  biLSTM Equalizer - Q-factor Metric}
\end{subfigure}
\hspace{\fill}
\begin{subfigure}[t]{0.5\textwidth}
            \centering
                \begin{tikzpicture}[scale=0.9]
      \begin{axis} [
        xlabel={Launch power [dBm]},
        ylabel={AIR [bits/symbol]},
        grid=both,   
    	xmin=-5, xmax=4,
    	xtick={-5, ..., 4},
        legend style={legend pos=south west, legend cell align=left,fill=white, fill opacity=0.6, draw opacity=1,text opacity=1},
    	grid style={dashed}]
        ]
           \addplot[color=red, mark=square, very thick]
    coordinates {
     (-5, 3.918) (-4, 3.937) (-3, 3.9455) (-2, 3.9543) (-1, 3.9670)
     (0, 3.9796) (1, 3.9819) (2, 3.9842) (3, 3.9783) (4, 3.9724)
    };
    \addlegendentry{\footnotesize{biLSTM Reg. (Test)}};

    \addplot[color=blue, mark=triangle, very thick]        coordinates {
     (-5, 3.908) (-4, 3.921) (-3, 3.9455) (-2, 3.9559) (-1, 3.9569)
     (0, 3.9579) (1, 3.9605) (2, 3.9632) (3, 3.9573) (4, 3.9513)
};
    \addlegendentry{\footnotesize{biLSTM MC Class. (Test)}};
    
               \addplot[color=red, mark=square,dashed, very thick]
    coordinates {
     (-5, 3.888) (-4, 3.9437) (-3, 3.9635) (-2, 3.9727) (-1, 3.9818)
     (0, 3.9863) (1, 3.9908) (2, 3.9908) (3, 3.9915) (4, 3.9922)

    };
    \addlegendentry{\footnotesize{biLSTM Reg. (Train)}};

    \addplot[color=blue, mark=triangle,dashed, very thick]        coordinates {
     (-5,  3.9692) (-4,  3.973) (-3,  3.978258586) (-2,  3.998990113) (-1,3.999422922)
     (0, 3.998268766) (1, 3.998268766) (2, 3.996826071) (3, 3.991632369) (4, 3.986438667
)
    };
    \addlegendentry{\footnotesize{biLSTM MC Class. (Train)}};

    \addplot[color=green, very thick]         coordinates {
     (-5, 3.888) (-4, 3.9137) (-3, 3.9155) (-2, 3.888) (-1, 3.8240)
     (0, 3.7097) (1, 3.5078) (2, 3.1945) (3, 2.79) (4, 2.2809)
    };
    \addlegendentry{\footnotesize{Regular DSP}};

    \end{axis}
    \end{tikzpicture}
    \caption{TWC link -  biLSTM Equalizer - AIR Metric}
\end{subfigure}
\begin{subfigure}[t]{0.5\textwidth}
            \centering 
              \begin{tikzpicture}[scale=0.9]
      \begin{axis} [
        xlabel={Launch power [dBm]},
        ylabel={Q-factor [dB]},
        grid=both,   
    	xmin=-5, xmax=4,
    	xtick={-5, ..., 4},
        legend style={legend pos=south west, legend cell align=left,fill=white, fill opacity=0.6, draw opacity=1,text opacity=1},
    	grid style={dashed}]
        ]
          \addplot[color=red, mark=square, very thick]
    coordinates {
     (-5, 7.63) (-4, 7.96) (-3, 8.21) (-2, 8.53) (-1, 8.65)
     (0, 8.56) (1, 8.43) (2, 8.26) (3, 7.86) (4, 7.22)
    };
    \addlegendentry{\footnotesize{MLP Reg. (Test)}};

    \addplot[color=blue, mark=triangle, very thick]        coordinates {
    (-5, 7.6) (-4,7.87) (-3, 8.05) (-2, 8.15) (-1, 8.08)
     (0, 8.01) (1, 7.85) (2, 7.62) (3, 7.24) (4, 6.6)
    };
    \addlegendentry{\footnotesize{MLP MC Class. (Test)}};
    
              \addplot[color=red, mark=square, dashed, very thick]
    coordinates {
     (-5, 7.7) (-4, 7.99) (-3, 8.25) (-2, 8.92) (-1, 8.97)
     (0, 8.84) (1, 9.17) (2, 8.38) (3, 8.29) (4, 7.54)
    };
    \addlegendentry{\footnotesize{MLP Reg. (Train)}};

    \addplot[color=blue, mark=triangle, dashed, very thick]        coordinates {
    (-5, 8.63) (-4,9.19) (-3, 9.46) (-2, 10) (-1, 10)
     (0, 9.8) (1, 9.59) (2, 9.30) (3, 8.9) (4, 7.7)
    };
    \addlegendentry{\footnotesize{MLP MC Class. (Train)}};
    
    \addplot[color=green, very thick]         coordinates {
   (-5, 7.83) (-4, 8.11) (-3, 8.25) (-2, 8.18) (-1, 7.8)
     (0, 7.17) (1, 6.33) (2, 5.24) (3, 3.97) (4, 2.48)
    };
    \addlegendentry{\footnotesize{Regular DSP}};

    \end{axis}
    \end{tikzpicture}
    \caption{TWC link -  MLP Equalizer - Q-factor Metric}
\end{subfigure}
\hspace{\fill}
\begin{subfigure}[t]{0.5\textwidth}
\centering 
              \begin{tikzpicture}[scale=0.9]
      \begin{axis} [
        xlabel={Launch power [dBm]},
        ylabel={AIR [bits/symbol]},
        grid=both,   
    	xmin=-5, xmax=4,
    	xtick={-5, ..., 4},
        legend style={legend pos=south west, legend cell align=left,fill=white, fill opacity=0.6, draw opacity=1,text opacity=1},
    	grid style={dashed}]
        ]
          \addplot[color=red, mark=square, very thick]
    coordinates {
     (-5, 3.888) (-4, 3.9137) (-3, 3.9155) (-2, 3.9282) (-1, 3.9180)
     (0, 3.9301) (1, 3.8999) (2, 3.8792) (3, 3.8417) (4, 3.7860)
 };
    \addlegendentry{\footnotesize{MLP Reg. (Test)}};

    \addplot[color=blue, mark=triangle, very thick]        coordinates {
     (-5, 3.888) (-4, 3.9137) (-3, 3.9155) (-2, 3.889770042
) (-1, 3.876072496
)
     (0, 3.870157446
) (1, 3.840149389
) (2, 3.819807389
) (3, 3.715211999
) (4, 3.592438651
)
    };
    \addlegendentry{\footnotesize{MLP MC Class. (Test)}};
    
              \addplot[color=red, mark=square, dashed, very thick]
    coordinates {
     (-5, 3.9018) (-4, 3.9337) (-3, 3.9555) (-2, 3.9751) (-1, 3.9708)
     (0, 3.9535) (1, 3.9508) (2, 3.9283) (3, 3.8775) (4, 3.8315)
    };
    \addlegendentry{\footnotesize{MLP Reg. (Train)}};

    \addplot[color=blue, mark=triangle, dashed, very thick]        coordinates {
     (-5,  3.9692) (-4,  3.973) (-3,  3.98493) (-2,  3.9993) (-1, 3.9996)
     (0, 3.9998) (1, 3.9988) (2, 3.9978) (3, 3.9942) (4, 3.9906)
    };
    \addlegendentry{\footnotesize{MLP MC Class. (Train)}};
    
    \addplot[color=green, very thick]         coordinates {
     (-5, 3.888) (-4, 3.9137) (-3, 3.9155) (-2, 3.888) (-1, 3.8240)
     (0, 3.7097) (1, 3.5078) (2, 3.1945) (3, 2.79) (4, 2.2809)
    };
    \addlegendentry{\footnotesize{Regular DSP}};

    \end{axis}
    \end{tikzpicture}
    \caption{TWC link -  MLP Equalizer - AIR Metric}
\end{subfigure}
\begin{subfigure}[t]{0.5\textwidth}
\centering 
            \begin{tikzpicture}[scale=0.9]
      \begin{axis} [
        xlabel={Launch power [dBm]},
        ylabel={Q-factor [dB]},
        grid=both,   
    	xmin=-3, xmax=5,
    	xtick={-3, ..., 5},
        legend style={legend pos=south west, legend cell align=left,fill=white, fill opacity=0.6, draw opacity=1,text opacity=1},
    	grid style={dashed}]
        ]
          \addplot[color=red, mark=square, very thick]
    coordinates {
         (-3,5.74)(-2,6.5)(-1,7.18)(0,8)(1,9)(2,9.7)(3,10.32)(4,10.62)(5,10.53)

    };
    \addlegendentry{\footnotesize{biLSTM Reg. (Test)}};

    \addplot[color=blue, mark=triangle, very thick]        coordinates {
    (-3,5.74)(-2,6.5)(-1,7.18)(0,8)(1,8.64)(2,8.79)(3,8.87)(4,9)(5,8.49)
    };
    \addlegendentry{\footnotesize{biLSTM MC Class. (Test)}};
    
              \addplot[color=red, mark=square,dashed, very thick]
    coordinates {
         (-3,5.9)(-2,6.7)(-1,7.2)(0,8.2)(1,9.3)(2,10)(3,10.64)(4,11)(5,11)

    };
    \addlegendentry{\footnotesize{biLSTM Reg. (Train)}};

    \addplot[color=blue, mark=triangle, dashed, very thick]        coordinates {
    (-3,6.15)(-2,7.8)(-1,7.5)(0,10)(1,13)(2,13)(3,13)(4,13)(5,13)
    };
    \addlegendentry{\footnotesize{biLSTM MC Class. (Train)}};

    \addplot[color=green, very thick]         coordinates {
       (-3,5.9)(-2,6.7)(-1,7.4)(0,8.2)(1,8.76)(2,9)(3,8.95)(4,8.49)(5,7.66)

    };
    \addlegendentry{\footnotesize{Regular DSP}};

    \end{axis}
    \end{tikzpicture}
    \caption{SSMF link - biLSTM Equalizer - Q-factor Metric}
\end{subfigure}
\hspace{\fill}
\begin{subfigure}[t]{0.5\textwidth}
            \centering 
            \begin{tikzpicture}[scale=0.9]
      \begin{axis} [
        xlabel={Launch power [dBm]},
        ylabel={AIR [bits/symbol]},
        grid=both,   
    	xmin=-3, xmax=5,
    	xtick={-3, ..., 5},
        legend style={legend pos=south west, legend cell align=left,fill=white, fill opacity=0.6, draw opacity=1,text opacity=1},
    	grid style={dashed}]
        ]
          \addplot[color=red, mark=square, very thick]
    coordinates {
         (-3,5.4718)(-2,5.6558)(-1,    5.8014)(0,5.8974)(1,5.9458)(2,5.9705)(3,5.9801)(4,5.9854)(5,
5.9870)
    };
    \addlegendentry{\footnotesize{biLSTM Reg (Test)}};

    \addplot[color=blue, mark=triangle, very thick]        coordinates {
         (-3,5.4718)(-2,5.6558)(-1,5.8104)(0,5.8920)(1,5.9392)(2,5.9464)(3,5.9486)(4,5.9498)(5,
5.9123)
    };
    \addlegendentry{\footnotesize{biLSTM MC Class. (Test)}};
    
              \addplot[color=red, mark=square,dashed, very thick]
    coordinates {
              (-3,5.4718)(-2,5.6558)(-1,5.8340)(0,5.9088)(1,5.9541)(2,5.9775)(3,5.9846)(4,5.9905)(5,
5.9898)
    };
    \addlegendentry{\footnotesize{biLSTM Reg. (Train)}};

    \addplot[color=blue, mark=triangle, dashed, very thick]        coordinates {
         (-3,5.5795)(-2,5.7895)(-1,5.8895)(0,5.9994)(1,5.9984)(2,5.9994)(3,5.9983)(4,5.9988)(5,
5.9988)
    };
    \addlegendentry{\footnotesize{biLSTM MC Class. (Train)}};

    \addplot[color=green, very thick]         coordinates {
         (-3,5.4718)(-2,5.6558)(-1,5.7895)(0,5.8794)(1,5.9266)(2,5.9444)(3,5.9422)(4,5.9032)(5,
5.8141)

    };
    \addlegendentry{\footnotesize{Regular DSP}};

    \end{axis}
    \end{tikzpicture}
    \caption{SSMF link - biLSTM Equalizer - AIR Metric}
\end{subfigure}
\end{figure*}
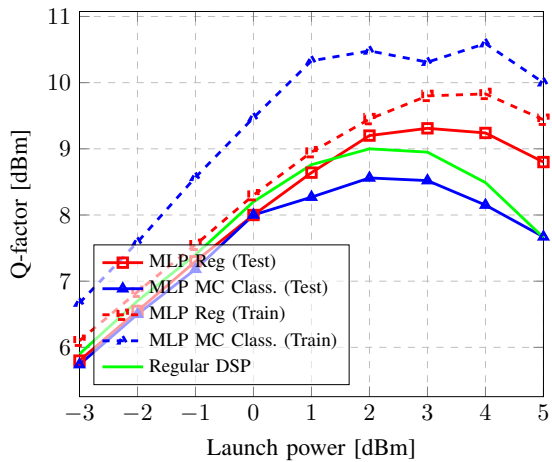
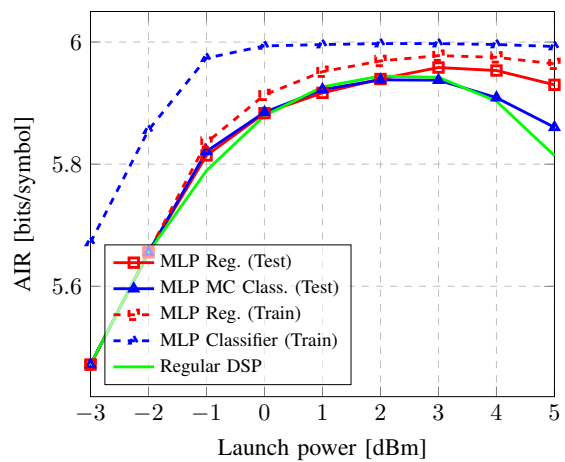
\begin{figure*}\ContinuedFloat
\begin{subfigure}[t]{0.5\textwidth}
            \centering 
              \begin{tikzpicture}[scale=0.9]
      \begin{axis} [
        xlabel={Launch power [dBm]},
        ylabel={Q-factor [dBm]},
        grid=both,   
    	xmin=-3, xmax=5,
    	xtick={-3, ..., 5},
        legend style={legend pos=south west, legend cell align=left,fill=white, fill opacity=0.6, draw opacity=1,text opacity=1},
    	grid style={dashed}]
        ]
          \addplot[color=red, mark=square, very thick]
    coordinates {
      (-3,5.8)(-2,6.55)(-1,7.3)(0,8)(1,8.64)(2,9.2)(3,9.31)(4,9.24)(5,8.8)
    };
    \addlegendentry{\footnotesize{MLP Reg (Test)}};

    \addplot[color=blue, mark=triangle, very thick]        coordinates {
     (-3,5.74)(-2,6.5)(-1,7.18)(0,8)(1,8.27)(2,8.56)(3,8.52)(4,8.15)(5,7.67)
    };
    \addlegendentry{\footnotesize{MLP MC Class. (Test)}};
    
              \addplot[color=red, mark=square,dashed, very thick]
    coordinates {
      (-3,6.1)(-2,6.84)(-1,7.55)(0,8.3)(1,8.95)(2,9.45)(3,9.8)(4,9.83)(5,9.44)
    };
    \addlegendentry{\footnotesize{MLP Reg (Train)}};

    \addplot[color=blue, mark=triangle,dashed, very thick]        coordinates {
     (-3,6.67)(-2,7.59)(-1,8.57)(0,9.47)(1,10.33)(2,10.48)(3,10.31)(4,10.59)(5,10)
    };
    \addlegendentry{\footnotesize{MLP MC Class. (Train)}};

    \addplot[color=green, very thick]         coordinates {
     (-3,5.9)(-2,6.7)(-1,7.4)(0,8.2)(1,8.76)(2,9)(3,8.95)(4,8.49)(5,7.66)
    };
    \addlegendentry{\footnotesize{Regular DSP}};

    \end{axis}
    \end{tikzpicture}
    \caption{SSMF link - MLP Equalizer - Q-factor Metric}
\end{subfigure}
\hspace{\fill}
\begin{subfigure}[t]{0.5\textwidth}
            \centering 
              \begin{tikzpicture}[scale=0.9]
      \begin{axis} [
        xlabel={Launch power [dBm]},
        ylabel={AIR [bits/symbol]},
        grid=both,   
    	xmin=-3, xmax=5,
    	xtick={-3, ..., 5},
        legend style={legend pos=south west, legend cell align=left,fill=white, fill opacity=0.6, draw opacity=1,text opacity=1},
    	grid style={dashed}]
        ]
          \addplot[color=red, mark=square, very thick]
    coordinates {
             (-3,5.4718)(-2,5.6558)(-1,5.8143)(0,5.8836)(1,5.9168)(2,5.9395)(3,5.9583)(4,5.9534)(5,
5.9300)

    };
    \addlegendentry{\footnotesize{MLP Reg. (Test)}};

    \addplot[color=blue, mark=triangle, very thick]        coordinates {
            (-3,5.4718)(-2,5.6558)(-1,5.8211)(0,5.8849)(1,5.9223)(2,5.9381)(3,5.9376)(4,5.9089)(5,
5.8604)

    };
    \addlegendentry{\footnotesize{MLP MC Class. (Test)}};
    
              \addplot[color=red, mark=square,dashed, very thick]
    coordinates {
              (-3,5.4718)(-2,5.6558)(-1,5.8354)(0,5.9135)(1,5.9515)(2,5.9691)(3,5.9778)(4,5.9750)(5,
5.9644)

    };
    \addlegendentry{\footnotesize{MLP Reg. (Train)}};

    \addplot[color=blue, mark=triangle,dashed, very thick]        coordinates {
             (-3,5.6718)(-2,5.8558)(-1,5.9737)(0,5.9937)(1,5.9959)(2,5.9975)(3,5.9975)(4,5.9960)(5,
5.9928)

    };
    \addlegendentry{\footnotesize{MLP Classifier (Train)}};

    \addplot[color=green, very thick]         coordinates {
           (-3,5.4718)(-2,5.6558)(-1,5.7895)(0,5.8794)(1,5.9266)(2,5.9444)(3,5.9422)(4,5.9032)(5,
5.8141)
    };
    \addlegendentry{\footnotesize{Regular DSP}};

    \end{axis}
    \end{tikzpicture}
    \caption{SSMF link - MLP Equalizer - AIR Metric}
\end{subfigure}
        \caption[ ]
        {Generalization study for regression equalizer (Reg.)  and Multi-Class Classifier (MC Class.)  showing the impact of overfitting in the NN performance and training process on the following scenarios: (a,b) biLSTM analyses and (c,d) MLP analyses for SC-DP-16-QAM, 9x50km TWC fiber and 34.4GBd; (e,f) biLSTM analyses and (g,h) MLP analyses for SC-DP-64-QAM, 5x100km SSMF fiber and 34.4GBd. } 
        \label{fig: overfitting}
\end{figure*}

\section{Comparison of Soft-Demapping based on Regression or Classification}

\subsection{Performance Comparison}

To test the importance of differentiating the output label misclassification, which takes place in the regression as opposed to the multi-class classification, we numerically simulated different modulation formats (16-QAM, 32-QAM, and 64-QAM)  transmissions of a single-channel (SC) DP 34.4~GBd signal with RRC 0.1 roll-off pulse over a system consisting of $5\!\times\!100$~km SSMF spans at 6 and 10~dBm launch power\footnote{Those transmission powers were picked since they provided BER different than zero in all three modulation formats and they are in the nonlinear regime}. In this first test, since only the modulation format changes, but we do not have any difference in nonlinearity strength, the goal is to show that when we increase the modulation format, the MC classifier's performance degrades because the categorical cross-entropy loss loses the information related to the miss-classification of different label types and values every misclassification occurrence equally. 

The comparison of regression-based and multi-class classification-based systems' performance for different modulation format orders is depicted in Fig.~\ref{figure:diffmod}. From the results of Fig.~\ref{figure:diffmod}~(a) and (c), one can see the impact on the MC Classifier's performance when increasing the number of classes in the problem (i.e. increasing the modulation format order), in terms of the Q-factor for 6  and 10 dBm, respectively.

For the biLSTM architecture, the percentage of how higher the Q factor is after regression equalization compared to the multi-class classification one was roughly 8\%, 31\%, and 31\% for the  16-QAM, 32-QAM, and 64-QAM scenarios, respectively, for the 6 dBm test case in Fig.~\ref{figure:diffmod}~(a), and 22\%, 29\%, and 50\% for the 16-QAM, 32-QAM, and 64-QAM scenarios, respectively, for the 10 dBm test case in Fig.~\ref{figure:diffmod}~(c).
For the MLP architecture, as compared to the multi-class classification output in  16-QAM, 32-QAM, and 64-QAM scenarios, the regression equalization always delivered better results, yielding 14\%, 18\%, and 15\% Q-factor improvement, respectively for the 6 dBm test case, Fig.~\ref{figure:diffmod}~(a), and 7\%, 11\%, and 19\% Q-factor improvement, respectively, for the 10 dBm test case, Fig.~\ref{figure:diffmod}~(c). When using the biLSTM layer, we can observe a greater difference between the regression and classification for different modulation formats, because the biLSTM layer, on average, performs much better than just MLP layers \cite{freire2021performance,9489894}. So, in the biLSTM case, we can see better how much the classification loss function gets degraded by ignoring the difference between distinct miss-classification occurrences.

When evaluating the performance in terms of AIR, almost the same behavior was observed: the results are depicted in Fig.~\ref{figure:diffmod} (b) and (d) for 6 and 10 dBm, respectively. In the case of the biLSTM architecture, the difference between the AIR obtained by regression and the multiclass classifier for 16-, 32-, and 64-QAM  was approximately  0.0008, 0.029, and 0.21, for 6~dBm; and  0.13, 0.33, and 0.63, for 10~dBm. Again, by increasing the order of the modulation format, the regression achieved better results than the multi-class classifiers\footnote{This trend was also observed in Ref.~\cite{end_to_end_MSE} in which they showed that the performance of CEL (classifiers) is worse when increasing the order of the QAM modulation format. In Ref.~\cite{end_to_end_MSE}, they concluded that more accurate gradients are needed for the case of the higher-order constellation, and with the sigmoid function and noisy gradient estimation, the BCEL failed to optimize the high-dimensional parameters.}.

In the case of 6~dBm with the MLP architecture, the same trend appeared again: the AIR difference was larger when the modulation order increased. The difference between AIR for regression vs. multi-class classification was 0.006, 0.057, and 0.21, for modulation formats 16-, 32-, and 64-QAM, respectively. However, for the 10dBm with MLP 64 QAM case (Fig.~\ref{figure:diffmod} (d)), the regression and multiclass classifier AIR bars are nearly identical. We believe that this can be interpreted as a learning limitation of the MLP architecture, and given that the true regression-based AIR value is underestimated, we can confidently state that the overall tendency discovered for all cases studied remains valid.

\begin{figure*}[b!]
   \subfloat[\label{genworkflow}]{%
      \includegraphics[width=0.32\textwidth]{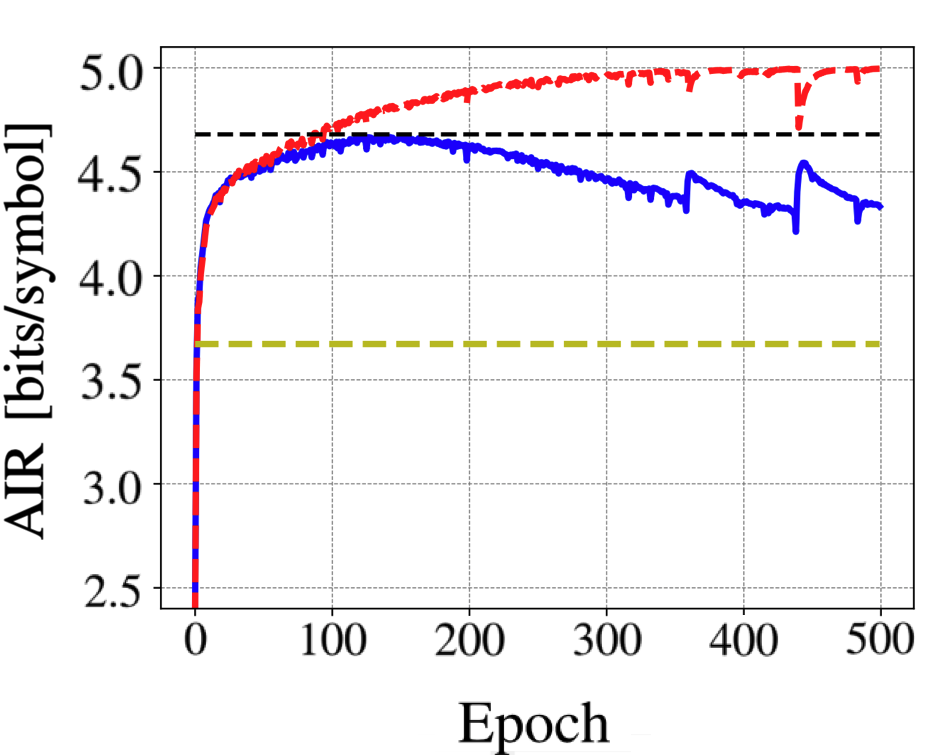}}
\hspace{\fill}
   \subfloat[\label{pyramidprocess} ]{%
      \includegraphics[width=0.32\textwidth]{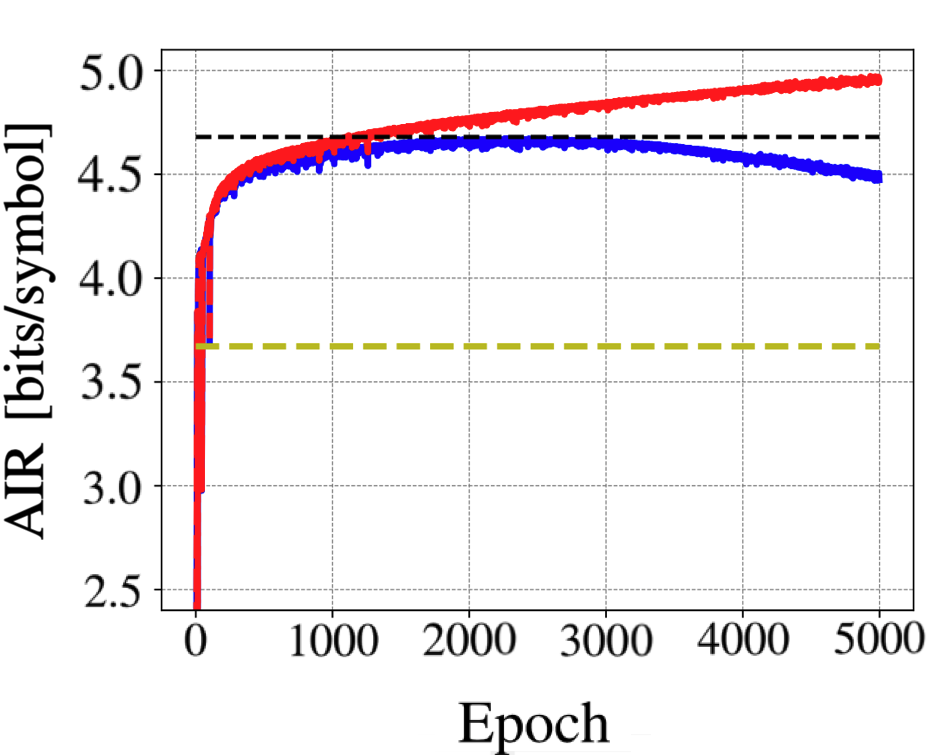}}
\hspace{\fill}
   \subfloat[\label{mt-simtask}]{%
      \includegraphics[width=0.32\textwidth]{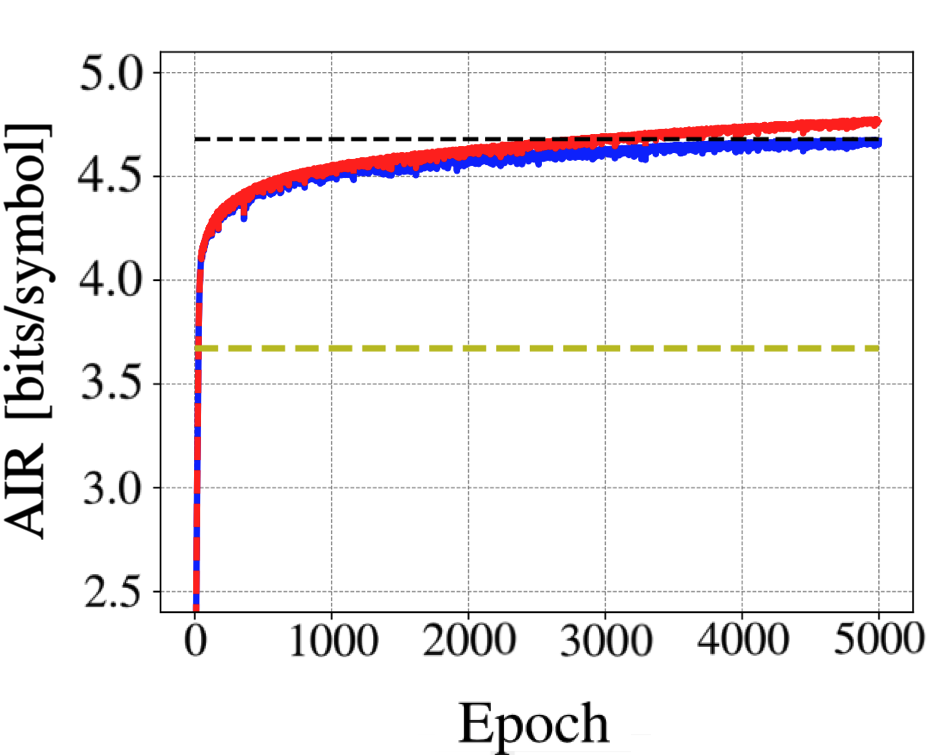}}\\
\caption{The performance (AIR) versus training epochs for the 32-QAM SSMF 10~dBm case using the biLSTM Multi-Class Classifier with learning rate equal to (a)  $10^{-3}$, (b) $ 10^{-4}$, and (c) $5\times 10^{-5}$. The red solid curve is the training performance, the blue solid curve is the test performance, the green dashed line is the reference performance when only the linear equalization is used, and the black dashed line is the reference for the maximum AIR achieved by the testing dataset over the training epochs. }
\label{Fig_train}
\end{figure*}

Now we turn to the issue of overfitting in the multi-class classification equalizer, addressing two cases: i) the case of a low dispersion fiber (SC-DP 16QAM 34.4 GBd over 9x50km TWC fiber), and ii) the case of a conventional SSMF fiber (SC-DP 64QAM 34.4 GBd over 5x100km SSMF fiber). To reveal the overfitting, for both the biLSTM and MLP equalizers, we present the Q-factor/AIR values for the training and test (validation) datasets. The difference between the values obtained in training and testing is the qualitative measure of the overfitting strength: the larger the difference, the stronger the overfitting. Therefore, the comparison of multi-class classification and regression training and testing results will reveal which approach generalizes better. Fig.~\ref{fig: overfitting} shows the results of our analysis where the solid green line is the Q-factor/AIR after only linear equalization (regular DSP), the solid blue and red lines indicate the Q-factor/AIR of the multi-class classification and regression models evaluated with the testing dataset, respectively,  and the dashed blue and red lines depict the Q-factor/AIR of the multi-class classification and regression models evaluated with the training dataset.

When we use the CEL in our equalizers, we see the same trend of higher overfitting levels as was observed in Ref.~\cite{bosman2020visualising}. As can be seen in all four panels, the Q-factor curves of training and testing for the multi-class classifiers show a significant difference (since this metric gives the logarithmic measure), suggesting the presence of noticeable overfitting in the classification model. Then, we can see that, in comparison to the multi-class classifier's result, the training and testing output curves, when using the regression, behave almost identically. It means that the regression model using MSE generalizes much better for all of our test cases (two different NN equalizers and two different transmission setups), which, again, complies with the conclusions reached in Ref.~\cite{bosman2020visualising}. Furthermore, we were able to see that, by using regression equalizers, the Q-factor level after equalization was still higher than with the classification, due to the better generalization of the regression NN models. When we look at the AIR values, we see that the multi-class classifier's training performance was overfitted, yielding virtually the maximum AIR attainable for each scenario, but in the case of regression, the training and testing curves followed the same trend, indicating a better generalization of the problem. Also, we notice that in the 64-QAM 5x100 SSMF transmission scenario, we have an even more reduced classification performance, with no increase in Q-factor observed for both biLSTM and MLP equalizers.

Finally, we would like to note that when we lowered the learning rate, we saw a reduction in overfitting in the classification task. Fig.~\ref{Fig_train} shows the example case of 32-QAM at 10~dBm using the SSMF link, where training and testing AIR curves for the biLSTM equalizer are shown for the three learning rates: $10^{-3}$, $ 10^{-4}$, and $5\times 10^{-5}$. As can be seen, for $10^{-3}$ the overfitting is much more intense than when using lower learning rates.  However, even with such low learning rates as $5 \times10^{-5}$,  the performance after 5000 epochs of training, was not better than that with either the regression or compared to the best case with the classification with $10^{-3}$ learning rate. Also, the overfitting could still be seen, as the training AIR level grew faster than the testing level. The maximum AIR measured with the test dataset is shown by the black dashed line in Fig.~\ref{Fig_train}. It is evident from this figure that lowering the learning rate did not result in any dramatic improvement in AIR for our test case. Next, we begin the investigation of the possible causes of the training challenges for the multi-class and multi-label classifiers using the learning rate equal to $ 10^{-4}$.

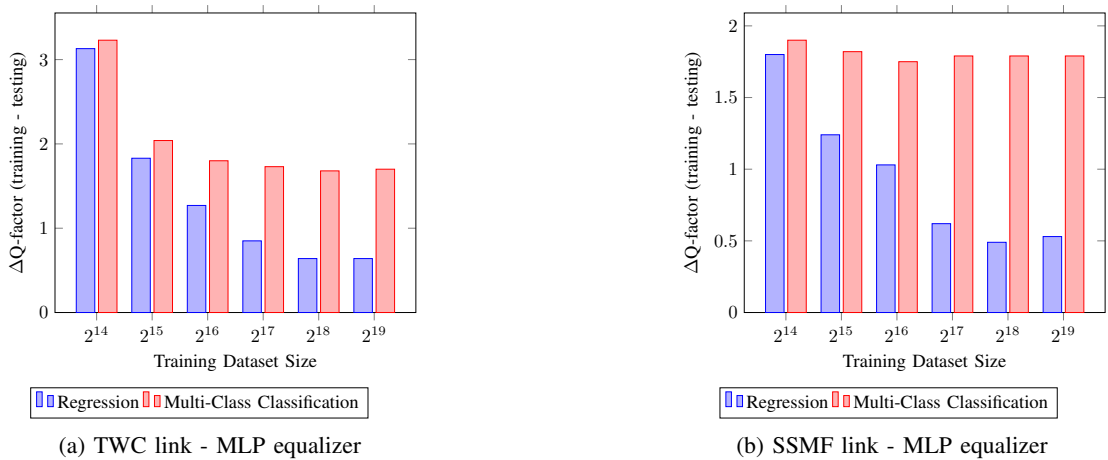
\begin{figure*}
\centering
\begin{subfigure}{.5\textwidth}
  \centering
    \begin{tikzpicture}[scale=0.7] 
\begin{axis}  
[  
    ybar, 
    enlarge x limits=0.15,
    legend style={at={(0.4,-0.25)}, anchor=north,legend columns=-1},     
     ylabel={ $\Delta$Q-factor (training - testing)}, 
     xlabel={Training Dataset Size}, 
    symbolic x coords={$2^{14}$, $2^{15}$, $2^{16}$, $2^{17}$, $2^{18}$, $2^{19}$}, xtick=data, 
    ymin=0,
    ]  
\addplot coordinates {($2^{14}$, 3.13) ($2^{15}$, 1.83) ($2^{16}$, 1.27)($2^{17}$, 0.85)($2^{18}$, 0.64)($2^{19}$, 0.64)}; %
\addplot coordinates {($2^{14}$, 3.23) ($2^{15}$, 2.04) ($2^{16}$, 1.8)($2^{17}$, 1.73)($2^{18}$, 1.68)($2^{19}$, 1.7)}; %
\legend{Regression, Multi-Class Classification}  
  
\end{axis}  
\end{tikzpicture}  
  \caption{TWC link - MLP equalizer}
  \label{fig:sub1}
\end{subfigure}%
\begin{subfigure}{.5\textwidth}
  \centering
    \begin{tikzpicture}[scale=0.7]  
\begin{axis}  
[  
    ybar, 
    enlarge x limits=0.15,
    legend style={at={(0.4,-0.25)}, anchor=north,legend columns=-1},     
     ylabel={ $\Delta$Q-factor (training - testing)}, 
     xlabel={Training Dataset Size}, 
    symbolic x coords={$2^{14}$, $2^{15}$, $2^{16}$, $2^{17}$, $2^{18}$, $2^{19}$},  
    xtick=data, 
    ymin=0,
    ]  
\addplot coordinates {($2^{14}$, 1.8) ($2^{15}$, 1.24) ($2^{16}$, 1.03)($2^{17}$, 0.62)($2^{18}$, 0.49)($2^{19}$, 0.53)}; %
\addplot coordinates {($2^{14}$, 1.9) ($2^{15}$, 1.82) ($2^{16}$, 1.75)($2^{17}$, 1.79)($2^{18}$, 1.79)($2^{19}$, 1.79)}; %
\legend{Regression, Multi-Class Classification}  
  
\end{axis}  
\end{tikzpicture}  
  \caption{SSMF link - MLP equalizer}
  \label{fig:sub2}
\end{subfigure}
\caption{Difference between training and testing performance over different training dataset sizes to evaluate the overfitting behavior. The MLP was applied for 3dBm 34.4GBd SC-DP, 9x50km TWC and 5x100km SSMF fiber links.}
\label{fig:overfitingdataset}
\end{figure*}

\subsection{Training Pitfalls and Overfitting Investigation}

In this final subsection, we outline the NN training process employed in this paper, taking into account not only the problem of overfitting but also the possible reasons why the classifiers struggled during learning and did not generalize well to the testing dataset. Once again, we emphasize that the CEL is the ideal loss function from the information theory viewpoint and that all of the performance degradation impacts we saw are directly attributable to the classifiers' gradient descent learning.

First, it is important to note that one possible criticism could be that the representation capability of the NN part may turn out to be insufficient to perform the demapping in the case of the classifier, while the regression could emerge as a ``simpler task'', such that the NN part is sufficient and does its job better in the latter case. Even though we do not discard that it might, potentially, play some role in specific cases (though it does not do in this current study), in this paper, we show that some fundamental machine learning-related issues underlie the worsening in the performance of classifiers as compared to the regression-based predictive modelling. Therefore, we claim that the NN`s capacity in the classification case is not the main problem, but the overfitting and gradient-related issues are the true cause. According to the literature [see, e.g., Ref.~\cite{Goodfellowbook2016}, Chapter 5.2], when an NN model is overfitting, our adding an extra complexity/capacity (say, in the form of additional hidden layers or neurons) would only worsen the overfitting effect. As a form of regularization, it is commonly suggested to do quite the opposite: to reduce the NN size for the sake of mitigating the strong overfitting effects. We have tested increasing the number of neurons in the classifier and observed a similar behaviour (for the same number of epochs) as we had had for the initial number of layers/weights, and Q-factor stayed around the ``original'' value. Additionally, we notice that even in the original case presented in Fig.~\ref{fig:NN_tasks}, the complexity of the regression-based equalizer and classifier is, in fact, not the same, despite their sharing the same NN part. In the regression, we have only two outputs, and in the MC classification, we have QAM-order-number neurons plus the softmax activation function that is responsible for the decision process, making the classification more complex than the regression in terms of multiplications number.

Having understood that the NN's capacity is not the problem's source, we now present the new results to show the true reason why the classifiers have high overfitting and demonstrate poorer performance in our study. The first and common hypothesis of such strong overfitting is that it occurs due to the lack of data (insufficient diversity) used in the training. To show that this is not the root cause of our problems, we have trained the NN classifier and regression equalizers, gradually enlarging the training data size from $2^{14}$ to $2^{19}$ symbols\footnote{For this particular study, the
data have not been randomly shuffled, and we use the same dataset size on all epochs}. In Fig.\ref{fig:overfitingdataset}, we show the difference in the Q factor between the training and testing runs, depending on the size of the dataset. As it is clear from this result, the strong overfitting was observed for the regression when training the system with a lower amount of data, and by increasing the dataset, the difference decreased to approximately 0.5 dB in both SSMF and TWC cases. The latter value can be reckoned as ``an acceptable margin'' to allow the system's generalization: training and testing give approximately the same result. However, when dealing with the classification, we see that increasing the dataset size is not enough: the difference between the training and testing results was still significant. Even though both regression and classification were trained with the same datasets, we can see that classification overfitting is much more pronounced, especially for the SSMF case.  Therefore, the important question raised in our work is: What can be the cause of this overfitting inclination in the classification case?

\begin{figure*}[htb]
    \centering 
\begin{subfigure}{0.33\textwidth}
  \includegraphics[width=\linewidth]{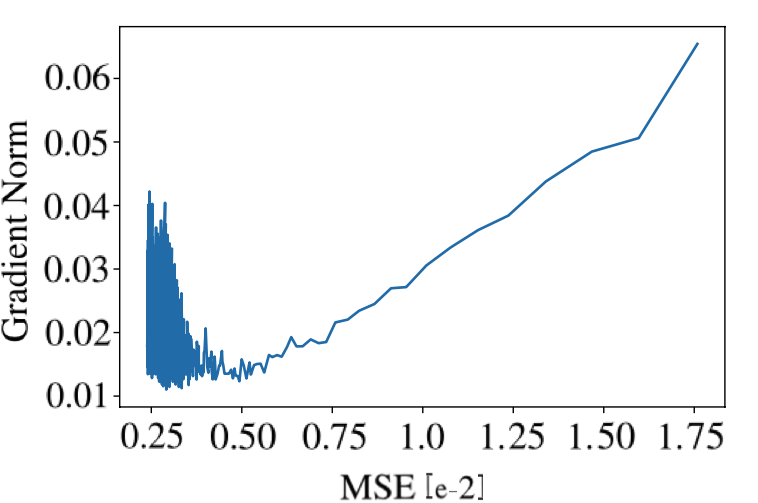}
  \caption{Total Grad Norm (Reg.)}
  \label{fig:1}
\end{subfigure}\hfil 
\begin{subfigure}{0.33\textwidth}
  \includegraphics[width=\linewidth]{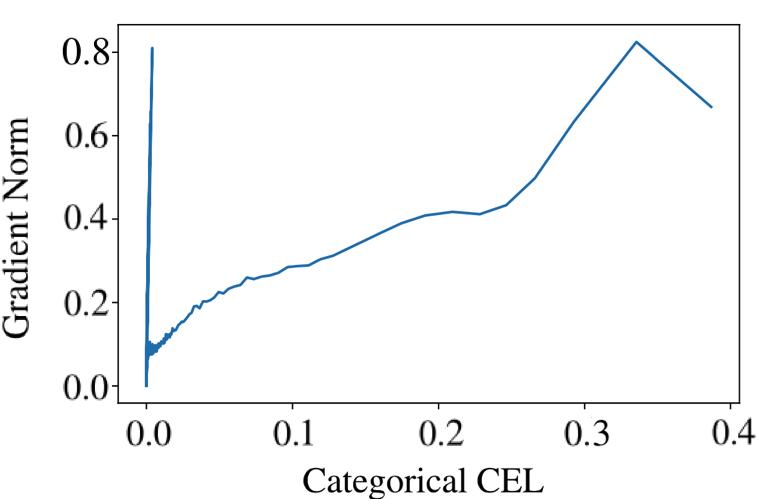}
  \caption{Total Grad Norm (MC Class.)}
  \label{fig:2}
\end{subfigure}\hfil 
\begin{subfigure}{0.33\textwidth}
  \includegraphics[width=\linewidth]{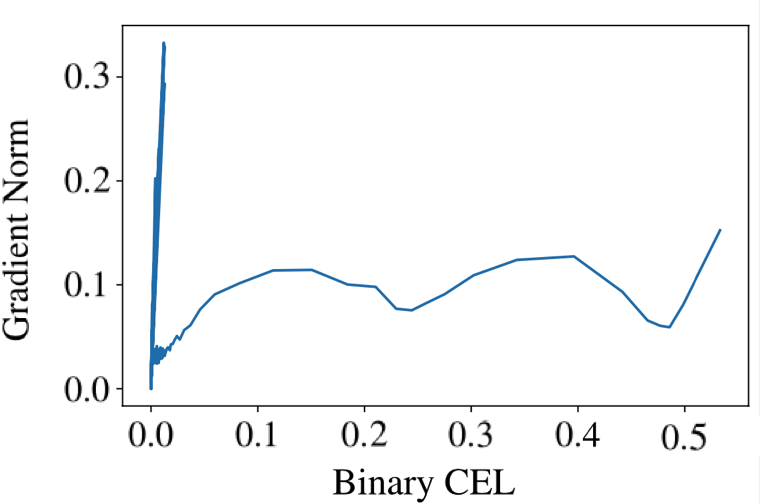}
  \caption{Total Grad Norm (ML Class.)}
  \label{fig:3}
\end{subfigure}
\caption{Gradient norm versus Training loss for the Regression model (a), Multi-Class Classifier (b), and Multi-Label Classifier (c) to investigate the gradient close to the minimum loss.}
\label{fig:gradnormevaluation}
\end{figure*}

\begin{figure*}[b!]
    \centering 
\begin{subfigure}{0.33\textwidth}
  \includegraphics[width=\linewidth]{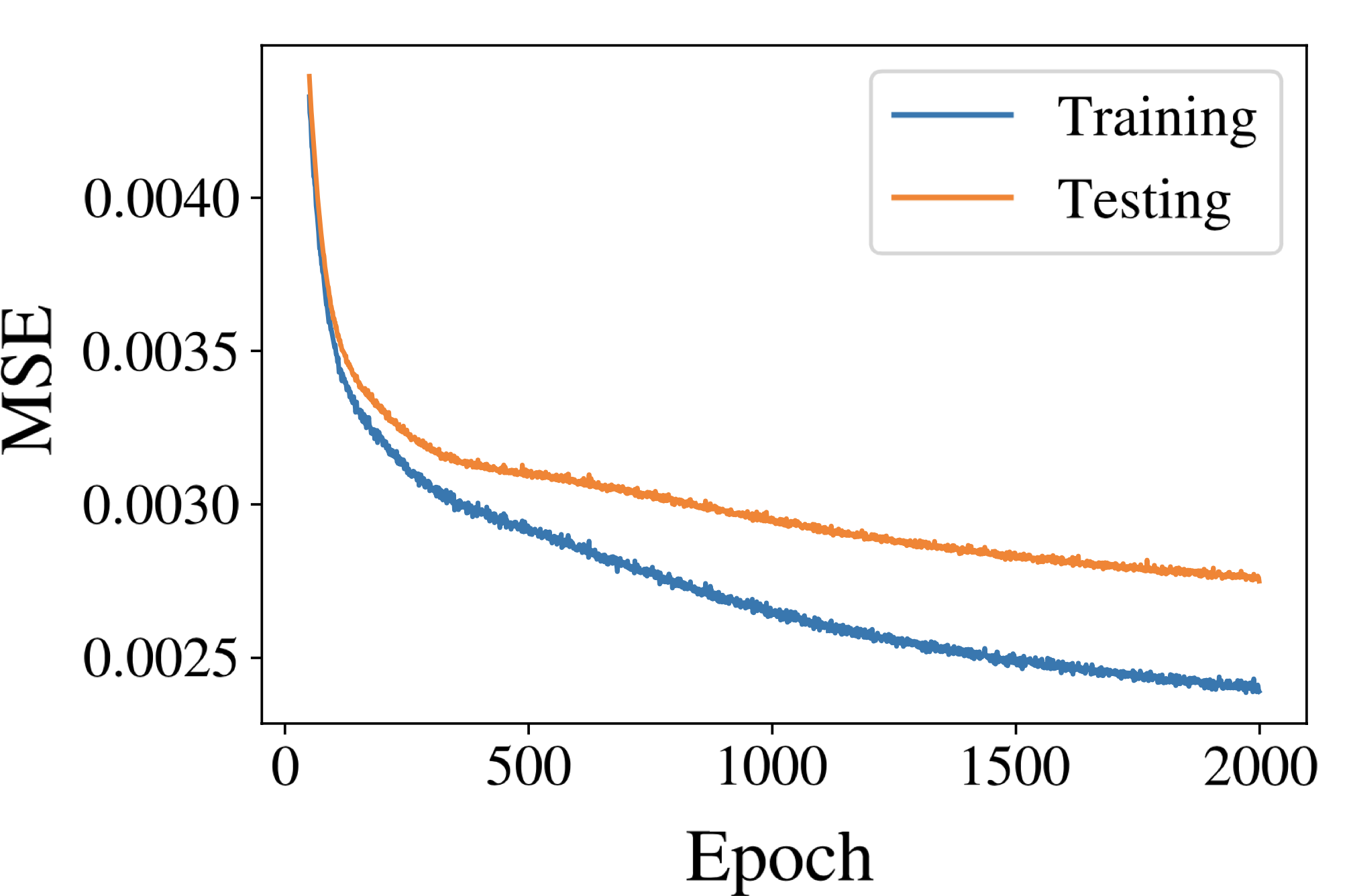}
  \caption{Loss vs. Epochs (Reg.)}
  \label{fig:4}
\end{subfigure}\hfil 
\begin{subfigure}{0.33\textwidth}
  \includegraphics[width=\linewidth]{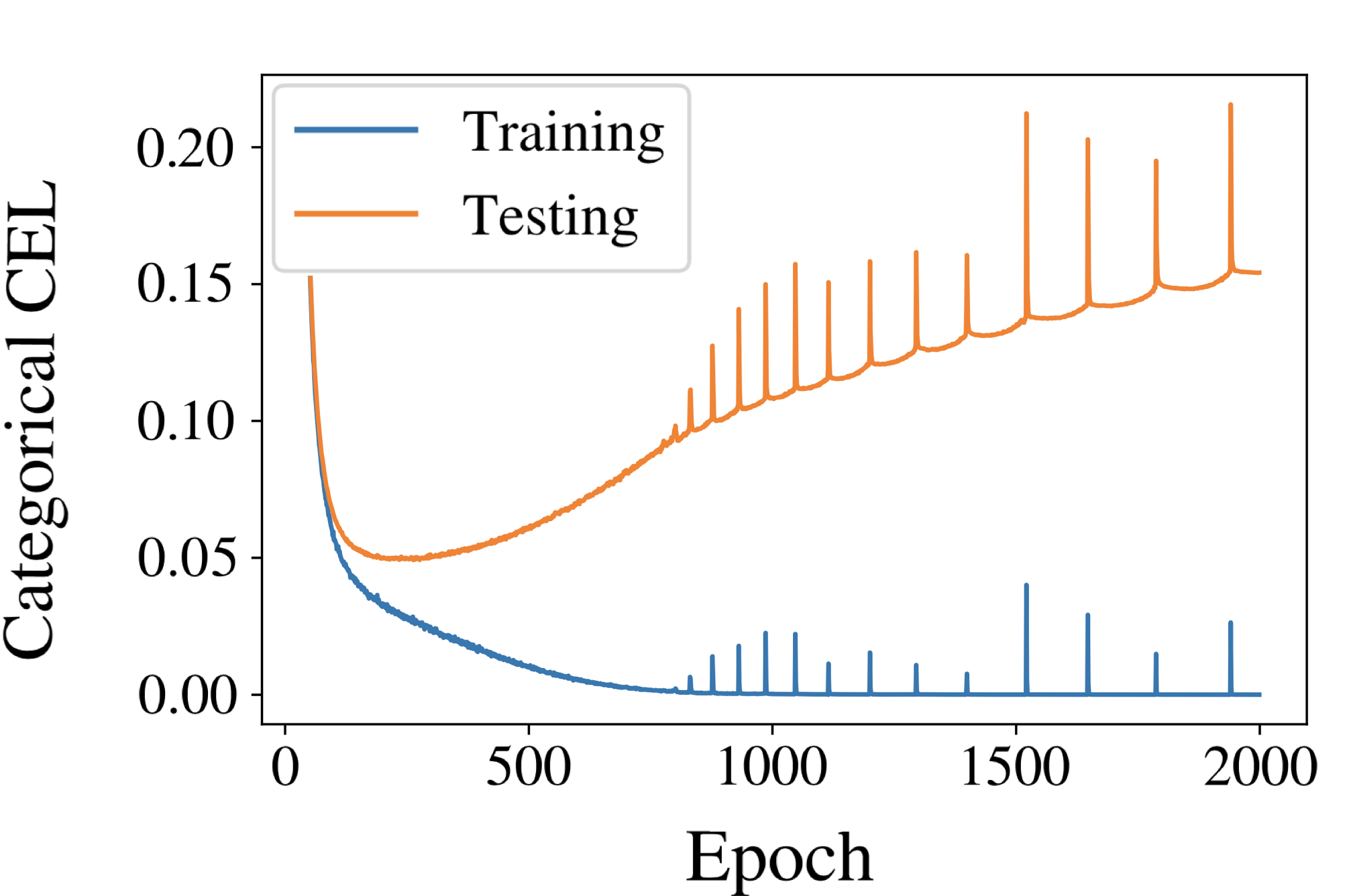}
  \caption{Loss vs. Epochs (MC Class.)}
  \label{fig:5}
\end{subfigure}\hfil 
\begin{subfigure}{0.33\textwidth}
  \includegraphics[width=\linewidth]{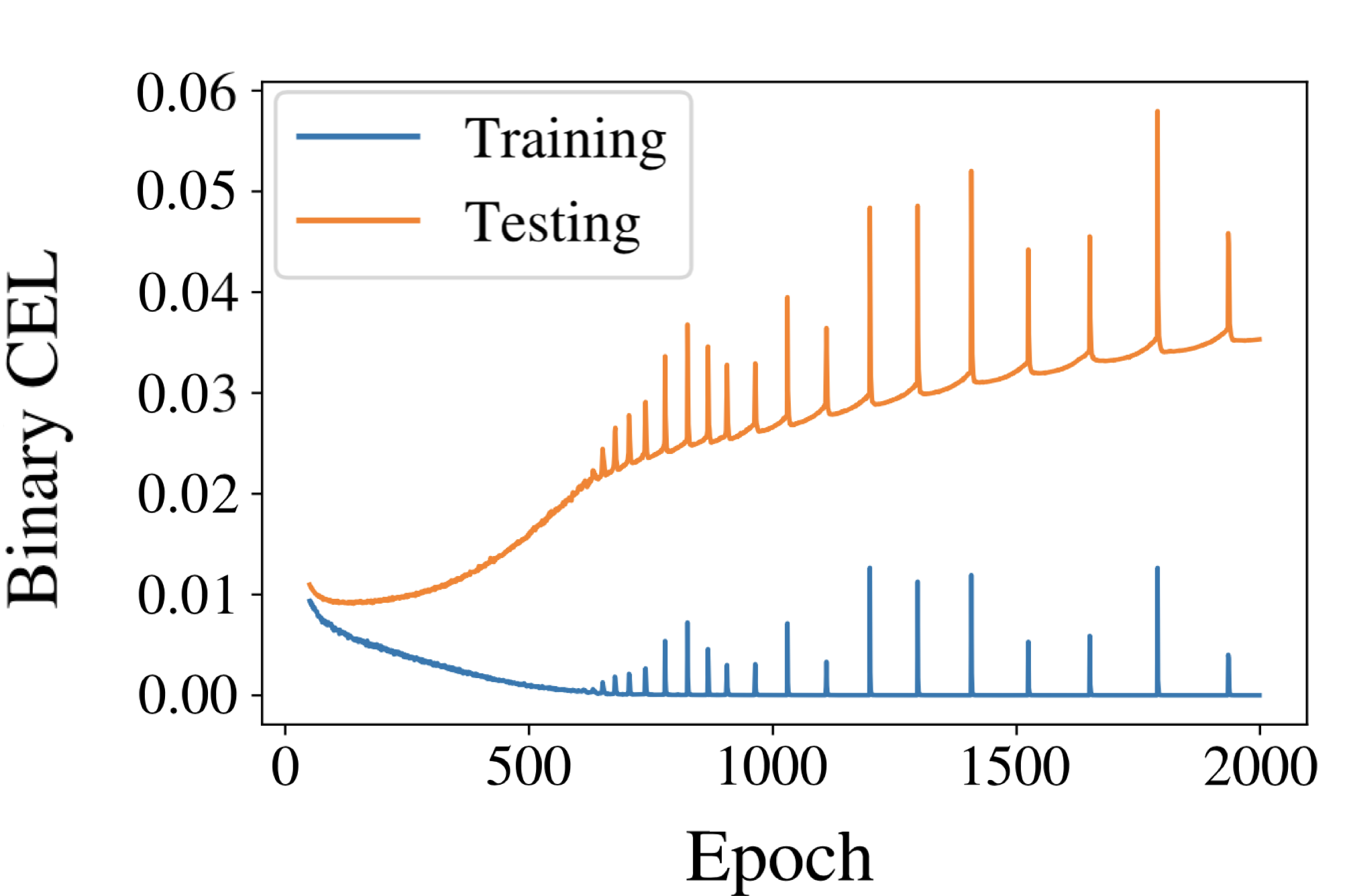}
  \caption{Loss vs. Epochs (ML Class.)}
  \label{fig:6}
\end{subfigure}
\medskip
\begin{subfigure}{0.33\textwidth}
  \includegraphics[width=\linewidth]{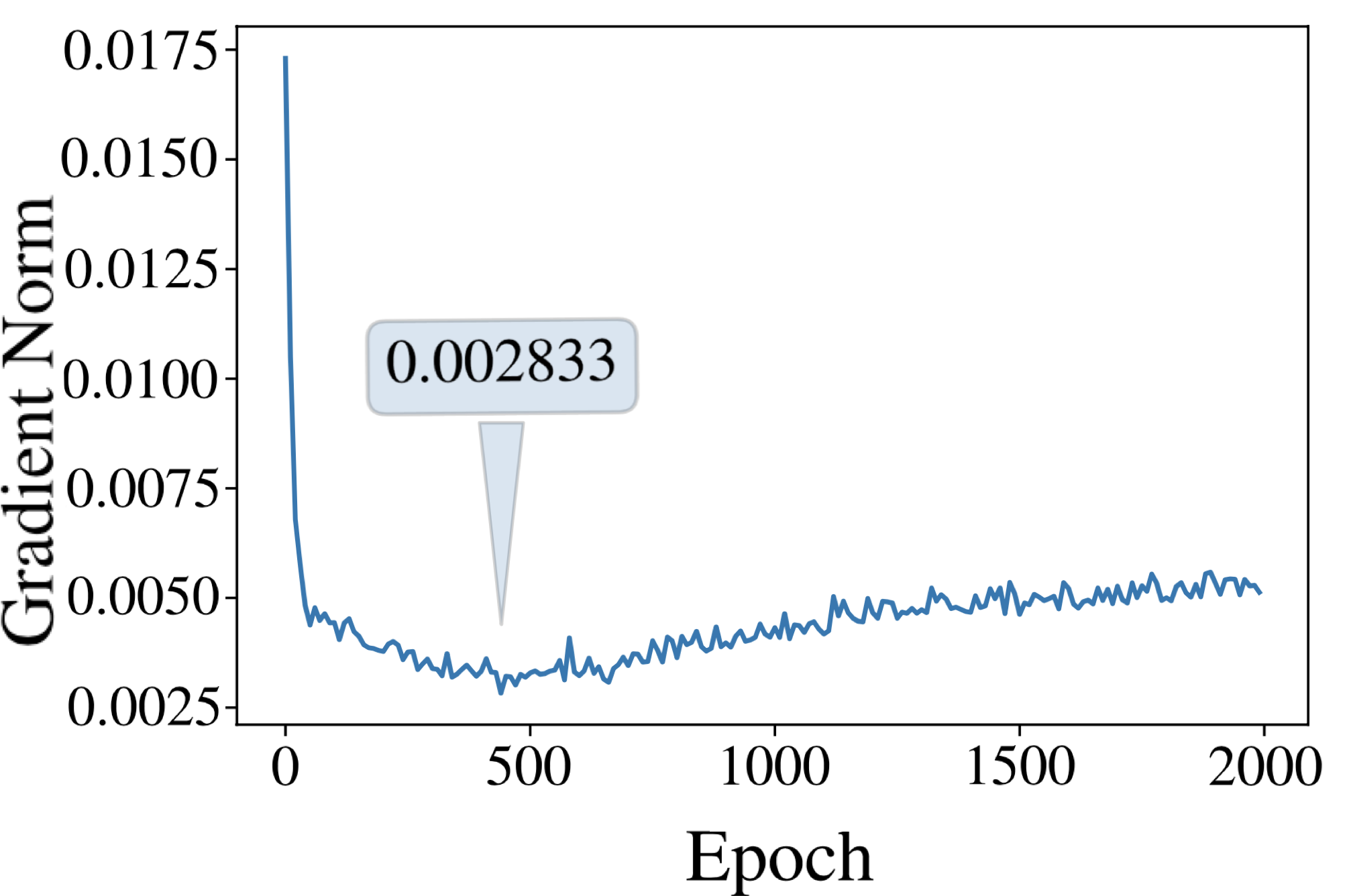}
  \caption{Grad Norm of $1^{o}$ layer (Reg.)}
  \label{fig:7}
\end{subfigure}\hfil 
\begin{subfigure}{0.33\textwidth}
  \includegraphics[width=\linewidth]{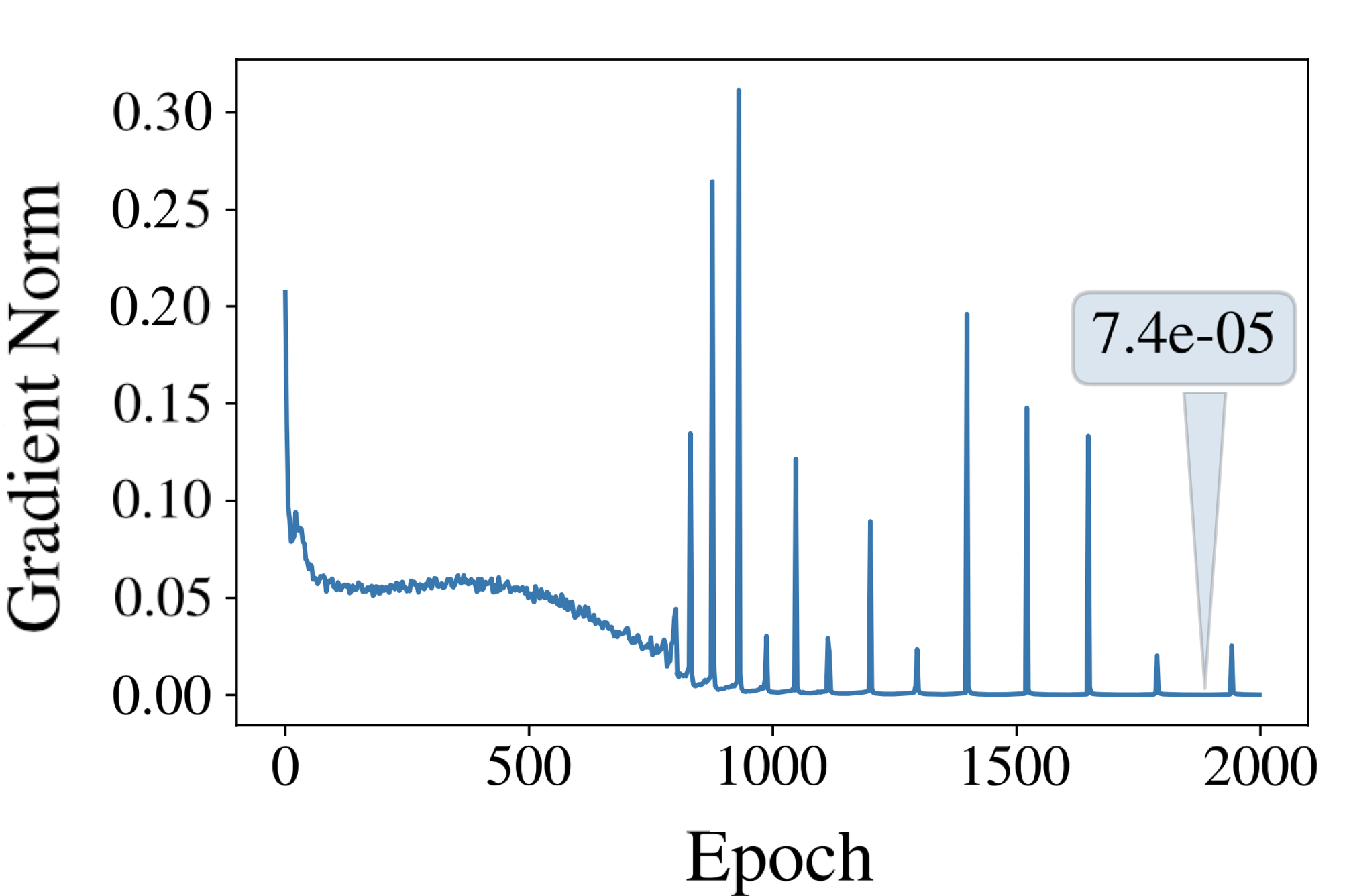}
  \caption{Grad Norm of $1^{o}$ layer (MC Class.)}
  \label{fig:8}
\end{subfigure}\hfil 
\begin{subfigure}{0.33\textwidth}
  \includegraphics[width=\linewidth]{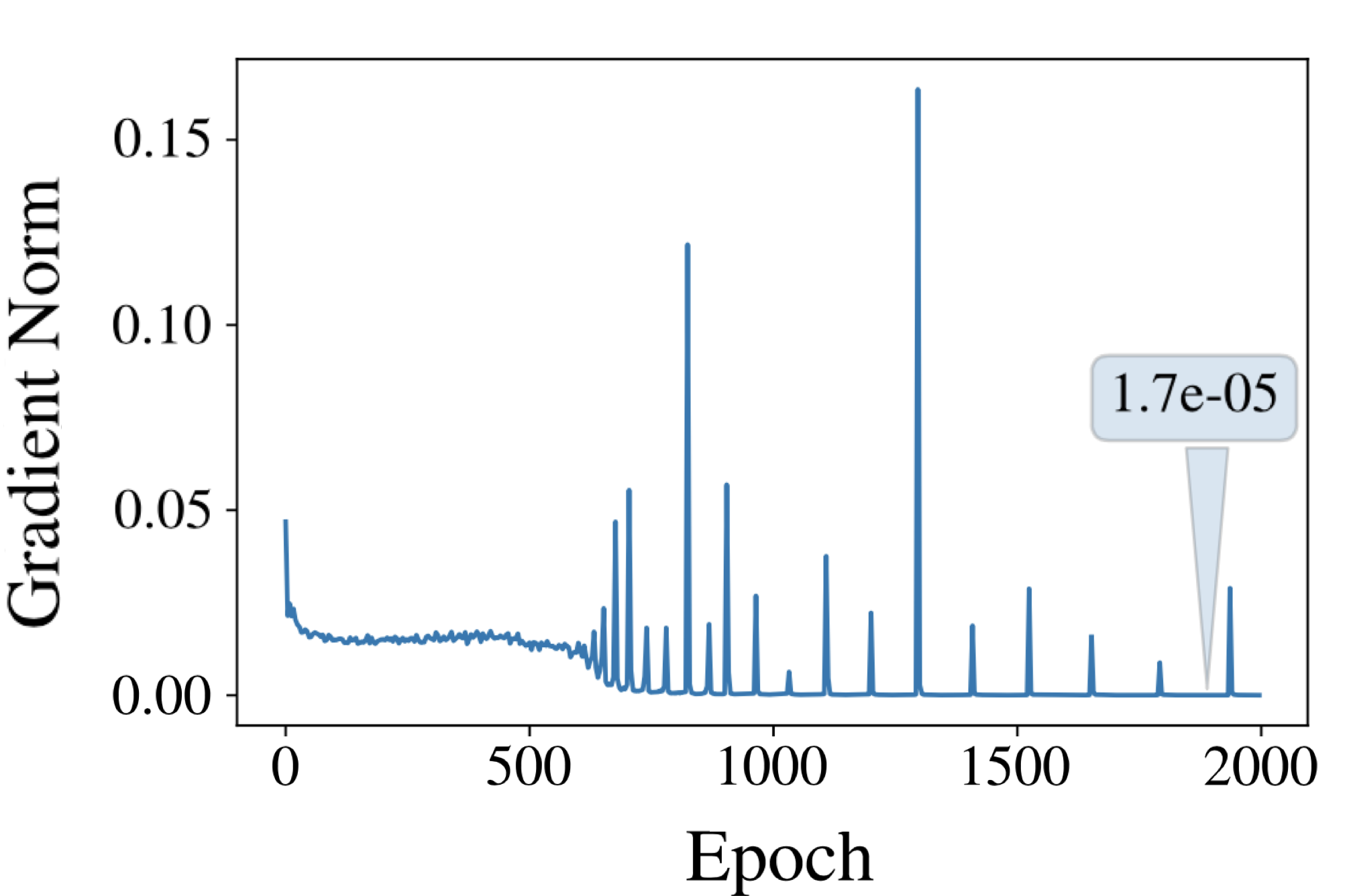}
  \caption{Grad Norm of $1^{o}$ layer (ML Class.)}
  \label{fig:9}
\end{subfigure}
\caption{Overfitting and Gradient Vanishing investigation for the Regression model (a/d), Multi-Class Classifier (MC Class.) (b/e), and Multi-Label Classifier (MB Class.) (c/f).}
\label{fig:vanishing}
\end{figure*}

Since, as we saw, the lack of data and the NN complexity/capacity, are not the main inducement for the classification system overfitting\footnote{Note that the overfitting problem can be rectified if a very large amount of training data is used or if we apply regularization procedures. However, the above measures do not address the gradient-related issues that are the genuine source of learning degradation for classifiers discussed in this section.}, we believe that the reason for such a behaviour is different. For the classification, the poor performance comes from the fact that for such high levels of accuracy as we have in the communications, the CEL produces really sharp local landscape minima that will cause the infamous problem of vanishing back-propagating gradients, making the classifier effectively cease learning. The fact of the sharp minima appearance in the CEL landscape can be shown by following the same method as in Ref.~\cite{bosman2020visualising}. To demonstrate this, we plot the gradient norm of the NN structure over the training loss value to evaluate whether a particular local minimum is sharp or wide. Our results are summarized in Fig.~\ref{fig:gradnormevaluation} ~(a), (b), and (c), for the regression, multi-class, and multi-label classification, respectively. Thus, in these figures, we recognize the same pattern as observed in Ref.~\cite[Fig. 15]{bosman2020visualising}: the CEL revealed multiple points of high gradient close to the loss minima (in our case the gradient norm reaches 0.8, and in Ref.~\cite{bosman2020visualising} it reached 0.5), and it is exactly the very indication of a sharp minimum in the loss landscape. On the other hand, the MSE was much less prone to overfitting due to weaker gradients close to the minima (in our case, those ranged from 0.01 to 0.04, in Ref.~\cite{bosman2020visualising} it reached 0.05). Therefore, our analysis demonstrates the sheer difference in the classification and regression cases: we have sharp local minima in the CEL landscape and do not have those for the MSE loss.

In addition to this analysis, a practical way to demonstrate that local minima cause the CEL learning problem is to verify that the gradient norm shrinks to some ``insignificant'' values along with the training, as described in Ref.~\cite[Chapter 8.2.2]{Goodfellowbook2016}. In Fig.~\ref{fig:vanishing}, we present the training and testing loss value over the epochs and the gradient norm of the first layer of each model as a function of the epoch number: the regression (MSE loss) in panels (a), (d),  multi-class -- in (b), (e),  multi-label --in panes (c), (f). In the lower row, we explicitly highlight the minimal gradient value over the epochs. In this gradient study, we consider only the MLP architecture applied to the SSMF fiber case. From this figure, we can clearly see that the gradient, when using the CEL, vanishes, whereas the same did not happen in the MSE loss case. When using the CCEL, the minimum gradient in the first layer reached $7.4e-5$, and when using the BCEL, it dropped further to $1.7e-5$; but, when using the MSE, the minimal gradient was around $0.003$. Here, we emphasize that the gradient value of $10^{-5}$ order is a clear indication of a vanishing gradient problem. Now, looking at the total gradient norm over the entire model (the absolute value of the sum of gradients for all MLP layers), we checked two cases: i) the gradient value at the epoch, where the testing loss value was lowest; ii) the gradient value after 2000 epochs.  For the MSE case, the gradient norm in i) was 0.017, and in ii) it was 0.010; in the case of CCEL, the gradient norm in i) was 0.083, and in ii) it was 1.17e-04.  Eventually, in the case of BCEL, the gradient norm in i) was 0.024, and in ii) it was 3.12e-05. With this result, we can see that the ``gradient shrinking to insignificant values along the training'' behaviour mentioned in Ref.~\cite{Goodfellowbook2016}, for both the CCEL and BCEL cases.  Also, we point out that the gradient vanishing problem in the softmax with CCEL was observed in Ref.~\cite{9662308}, and the same for the sigmoid with the BCEL, was reported in Ref.~\cite{end_to_end_MSE}. We claim that the same phenomenon happens in communications due to the high accuracy (low error) levels in input datasets that we have to work with, stipulating the difference in the performance of classification and regression-based predictive modelling that we report here.

Additionally, for this same transmission case, Fig.~\ref{fig:vanishing}, we could see that the original MSE before the NN, was 0.0066, and the CCEL before the NN, assuming the Gaussian distributions of constellation points, was 0.057. After the NN, the MSE of the regression was reduced to 0.0027 ($59\%$ drop in loss), and the CCEL of the multi-class classifier fell down to 0.049 ($15\%$ drop in loss). This fact, together with the gradient analyses, shows that for such high accuracy systems, where at the beginning of the training the loss function value is already very small, the MSE can still avoid vanishing of gradients, while the CEL falls in really sharp local minima and stops learning due to the vanishing gradient problem. Furthermore, we tried to address the overfitting issue with traditional methods (increasing the training dataset size, adding a dropout layer, batch normalization layer, and introducing a regularizer). However, even though the overfitting was partially mitigated, the Q-factor on a testing dataset was still almost the same because the vanishing gradients issue had not been solved with those measures.

\begin{figure}[ht!]
    \centering
    \begin{tikzpicture}[scale=0.85]
      \begin{axis}[
        xlabel={QAM Order},
        ylabel={Q-factor [dB] on the Testing Dataset.},
        set layers,
        ybar=1.2pt,
        bar width=6pt,
        symbolic x coords={16,32,64},
        grid=both,
        ymin=0,
        samples=4,
        legend cell align=left,
        ymax=13,
        x label style={font=\footnotesize},
        y label style={font=\footnotesize},
        ticklabel style={font=\footnotesize},
        xtick={16,32,64},
        legend style={at={(0.5,-0.3)}, anchor=north,legend columns=-1},
        ]
        
        \addplot[black,fill=red] coordinates {
      (16, 12.16) (32, 10.757) (64, 8.1857) 
        };
        \addlegendentry{Reg.}
                \addplot[black,fill=orange] coordinates {
        (16,10.9547) (32, 9.1506) (64, 7.1032) 
        };
        \addlegendentry{MC Class.}
        
        \addplot[black,fill=blue] coordinates {
        (16, 10.85) (32, 8.77) (64,6.94) 
        };
        \addlegendentry{ML Class.}
        \addplot[black,fill=green] coordinates {
         (16, 10.85) (32, 8.77) (64, 6.54) 
        };
        \addlegendentry{DSP Ref.}
      \end{axis}
    \end{tikzpicture}
    \caption{Performance evaluation on the testing datasets for three different modulation formats (16-,32-, and 64-QAM). The MLP was applied for 6dBm SC-DP, 5x100km 34.4GBd SSMF fiber links.}
    \label{fig:comparison_BCEL}
\end{figure}
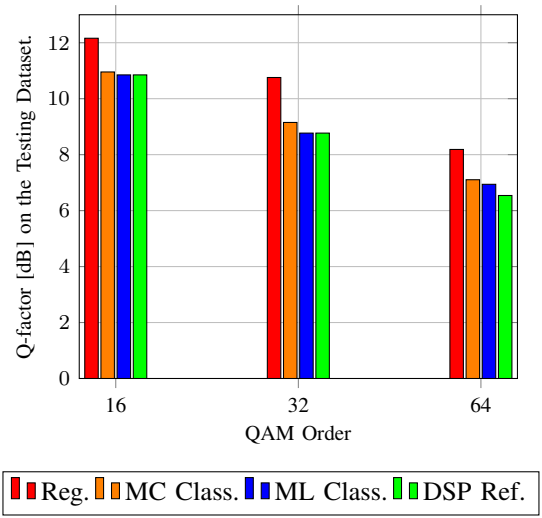

Finally, we would like to show the Q-factor performance of the multi-label classification using the BCEL. As we could see in Figs.~\ref{fig:gradnormevaluation}, \ref{fig:vanishing} the BCEL plus sigmoid suffered  from the same problems as the CCEL plus a softmax layer, and this translated into the degradation of performance as well. In terms of performance, when the MLP with a learning rate of 0.0001, was applied to the 64QAM, 3dBm, SC-DP, 5$\times$100~km 34.4~GBd SSMF link, the maximal Q-factor achieved by the regression was 9.6~dB, for the multi-class classifier it was 8.8dB, and for the multi-label classifier it was just 8.7 dB, showing that the change in the loss function and the addition of one extra layer did not help improve BER. Also, we appended the results in Fig \ref{figure:diffmod} a) with those referring to the multi-label classification: it is depicted in Fig. \ref{fig:comparison_BCEL}. There, we can see a trend similar to that reported in Ref.~\cite{end_to_end_MSE}: the regression equalization always delivered better results, yielding 14\%, 18\%, and 15\% Q-factor improvement for 16-QAM, 32-QAM, and 64-QAM, respectively, compared to the multi-class classification output. Turning to the BCEL case (multi-label classifier), this difference becomes even greater: the regression equalization improved the Q-factor by 15\%, 23\%, and 18\%, for the 16-QAM, 32-QAM, and 64-QAM scenarios, respectively.

We conclude this section with a comparison to a typical computer vision challenge and our view on the next steps in studying the loss function landscape occurring in communications-related applications. In computer vision, it was discovered that when an image is processed, the bulk of the pixies describes a graphic background, and only a few pixels pertain to the genuine objects in the image. This resulted in inefficient training because most parts of the image correspond to ``an easy prediction'' (which means that they can be easily labeled as background by the detector) and therefore offer little relevant learning. Although they individually provide tiny contributions to the loss value, when we combine those contributions, they can overwhelm the loss and computed gradients, resulting in a degraded model's prediction performance, since easy predictions (detections with high probabilities, or, in our context, the correct classifications following a simple hard decision) account for a large share of inputs. To address this issue, in Ref.~\cite{lin2017focal} Facebook A. I. developed a new modified approach named focal loss (FL), by adding a weighting factor to the cross-entropy loss. The FL gives a higher weight to cases that are hardly misclassified: in communications, it would correspond to the cases that got misclassified after the hard-decision (HD) process.
We believe that the difficulty of the ``dataset imbalance'' (meaning that just a small fraction of the dataset corresponds to the wrong HD predictions) exists in the high-accuracy communication-related equalization/demapping problem. As a toy example, consider
a system where an initial SER after the HD, is equal to $10^{-3}$. Training the NN-classifier, in this case,
will mean that a 99.9\% fraction of the training dataset corresponds to
``an easy prediction'', and the remaining 0.1\% will be the ``hard prediction'' members. Therefore, to improve the performance of classifiers, we expect that a similar focal loss function, as in computer vision, must be created for the communications application.

\section{Conclusion}

In this work, we compared the performance and training peculiarities of the regression and classification predictive models, addressing the NN-based
soft-demapping in coherent optical communication.
We considered several transmission scenarios, including three different modulation formats, on two different optical link test benches with different nonlinear and dispersion responses.
The applied NN models were based on two different architectures: the feedforward and recurrent NNs. For the regression equalizer and the multi-class classifier, the model had the same structure, except for the configuration of the last layer conditioned by the particular predictive modeling tasks.
In most of the scenarios studied in this work, the soft-demapping based on regression 
outperforms the one based on classification providing higher Q-factor and achievable information rate. We have further observed that 
the soft-demapping based on cross-entropy learning required was more prone to overfitting than the regression-based counterparts. 
This observation regarding overfitting is in line with findings from Ref.~\cite{bosman2020visualising} showing the performance advantage of regression models over classification models in a different context, due to the better generalization capability of the former. 

We should emphasize that both the regression and classification tasks have certain limitations. The regression loss function (the MSE) is a special case of the classification loss function (the CEL), in which the stochastic component of the output variables is assumed to be signal-independent and normally distributed. Therefore, the MSE does not take into account the signal-dependent stochastic contribution, which is, obviously, present in the true nonlinear optical channel.

Nevertheless, we underline that from the machine learning methods' application perspective, the classification loss function (the CEL) landscape typically involves very sharp local minima, which can cause the NN model to generalize much worse than the regression model with the loss based on the Euclidean distance. Additionally, we show that due to the high accuracy transmissions involved in communications, the CEL tends to vanish the gradients while the MSE can still maintain a staple gradient landscape that makes the NN learn more effectively.

Finally, although we have observed a common performance trend between these two predictive models, a general conclusion cannot be made. For instance, some advanced problem-specific loss functions or data pre-processing/mining may potentially hinder the training problems in the classification case. However, the scope of this work was to evaluate the performance of the typical classifier and regression architectures with their conventional loss function and we leave it to future investigation on how to improve the training process on NN classifiers in the task of soft demapping.

\bibliographystyle{IEEEtran}
\bibliography{references}

\end{document}